\documentclass[twocolumn]{aa}

\usepackage{tikz}
\usetikzlibrary{trees}
\usepackage{listings}
\usepackage{graphicx}
\usepackage{amsmath,amsfonts,amssymb}
\usepackage{txfonts}
\usepackage{color}
\usepackage{natbib}
\usepackage{float}
\usepackage{csquotes}
\usepackage{dblfloatfix}
\usepackage{afterpage}
\usepackage{ifthen}
\usepackage[morefloats=12]{morefloats}
\usepackage{placeins}
\usepackage{multicol}
\usepackage{makecell}
\usepackage{amsmath}
\usepackage{mathtools}
\bibpunct{(}{)}{;}{a}{}{,}
\usepackage[switch]{lineno}
\definecolor{linkcolor}{rgb}{0.6,0,0}
\definecolor{citecolor}{rgb}{0,0,0.75}
\definecolor{urlcolor}{rgb}{0.12,0.46,0.7}
\PassOptionsToPackage{hyphens}{url}
\usepackage[breaklinks, colorlinks, urlcolor=urlcolor,
    linkcolor=linkcolor,citecolor=citecolor,pdfencoding=auto]{hyperref}
\hypersetup{linktocpage}
\usepackage[most]{tcolorbox}

\usepackage{float}
\usepackage{listings}

\newfloat{lstfloat}{htbp}{lop}
\floatname{lstfloat}{Listing}

\def\setsymbol#1#2{\expandafter\def\csname #1\endcsname{#2}}
\def\getsymbol#1{\csname #1\endcsname}

\def\Planck{\textit{Planck}}

\newbox\tablebox    \newdimen\tablewidth
\def\leaderfil{\leaders\hbox to 5pt{\hss.\hss}\hfil}
\def\endPlancktablewide{\tablewidth=\textwidth 
    $$\hss\copy\tablebox\hss$$
    \vskip-\lastskip\vskip -2pt}
\def\tablenote#1 #2\par{\begingroup \parindent=0.8em
    \abovedisplayshortskip=0pt\belowdisplayshortskip=0pt
    \noindent
    $$\hss\vbox{\hsize\tablewidth \hangindent=\parindent \hangafter=1 \noindent
    \hbox to \parindent{$^#1$\hss}\strut#2\strut\par}\hss$$
    \endgroup}
\def\doubleline{\vskip 3pt\hrule \vskip 1.5pt \hrule \vskip 5pt}

\def\L2{\ifmmode L_2\else $L_2$\fi}

\def\DeltaT{\ifmmode \Delta T\else $\Delta T$\fi}
\def\deltat{\ifmmode \Delta t\else $\Delta t$\fi}
\def\fknee{\ifmmode f_{\rm knee}\else $f_{\rm knee}$\fi}
\def\Fmax{\ifmmode F_{\rm max}\else $F_{\rm max}$\fi}
\def\solar{\ifmmode{\rm M}_{\mathord\odot}\else${\rm M}_{\mathord\odot}$\fi}
\def\Msolar{\ifmmode{\rm M}_{\mathord\odot}\else${\rm M}_{\mathord\odot}$\fi}
\def\Lsolar{\ifmmode{\rm L}_{\mathord\odot}\else${\rm L}_{\mathord\odot}$\fi}
\def\inv{\ifmmode^{-1}\else$^{-1}$\fi}
\def\mo{\ifmmode^{-1}\else$^{-1}$\fi}
\def\sup#1{\ifmmode ^{\rm #1}\else $^{\rm #1}$\fi}
\def\expo#1{\ifmmode \times 10^{#1}\else $\times 10^{#1}$\fi}
\def\,{\thinspace}
\def\lsim{\mathrel{\raise .4ex\hbox{\rlap{$<$}\lower 1.2ex\hbox{$\sim$}}}}
\def\gsim{\mathrel{\raise .4ex\hbox{\rlap{$>$}\lower 1.2ex\hbox{$\sim$}}}}

\def\simprop{\mathrel{\raise .4ex\hbox{\rlap{$\propto$}\lower 1.2ex\hbox{$\sim$}}}}
\def\deg{\ifmmode^\circ\else$^\circ$\fi}
\def\pdeg{\ifmmode $\setbox0=\hbox{$^{\circ}$}\rlap{\hskip.11\wd0 .}$^{\circ}
          \else \setbox0=\hbox{$^{\circ}$}\rlap{\hskip.11\wd0 .}$^{\circ}$\fi}
\def\arcs{\ifmmode {^{\scriptstyle\prime\prime}}
          \else $^{\scriptstyle\prime\prime}$\fi}
\def\arcm{\ifmmode {^{\scriptstyle\prime}}
          \else $^{\scriptstyle\prime}$\fi}
\newdimen\sa  \newdimen\sb
\def\parcs{\sa=.07em \sb=.03em
     \ifmmode \hbox{\rlap{.}}^{\scriptstyle\prime\kern -\sb\prime}\hbox{\kern -\sa}
     \else \rlap{.}$^{\scriptstyle\prime\kern -\sb\prime}$\kern -\sa\fi}
\def\parcm{\sa=.08em \sb=.03em
     \ifmmode \hbox{\rlap{.}\kern\sa}^{\scriptstyle\prime}\hbox{\kern-\sb}
     \else \rlap{.}\kern\sa$^{\scriptstyle\prime}$\kern-\sb\fi}
\def\ra[#1 #2 #3.#4]{#1\sup{h}#2\sup{m}#3\sup{s}\llap.#4}
\def\dec[#1 #2 #3.#4]{#1\deg#2\arcm#3\arcs\llap.#4}
\def\deco[#1 #2 #3]{#1\deg#2\arcm#3\arcs}
\def\rra[#1 #2]{#1\sup{h}#2\sup{m}}

\def\dots{\relax\ifmmode \ldots\else $\ldots$\fi}
\def\WHzsr{\ifmmode $W\,Hz\mo\,sr\mo$\else W\,Hz\mo\,sr\mo\fi}
\def\mHz{\ifmmode $\,mHz$\else \,mHz\fi}
\def\GHz{\ifmmode $\,GHz$\else \,GHz\fi}
\def\mKs{\ifmmode $\,mK\,s$^{1/2}\else \,mK\,s$^{1/2}$\fi}
\def\muKs{\ifmmode \,\mu$K\,s$^{1/2}\else \,$\mu$K\,s$^{1/2}$\fi}
\def\muKRJs{\ifmmode \,\mu$K$_{\rm RJ}$\,s$^{1/2}\else \,$\mu$K$_{\rm RJ}$\,s$^{1/2}$\fi}
\def\muKHz{\ifmmode \,\mu$K\,Hz$^{-1/2}\else \,$\mu$K\,Hz$^{-1/2}$\fi}
\def\MJysr{\ifmmode \,$MJy\,sr\mo$\else \,MJy\,sr\mo\fi}
\def\MJysrmK{\ifmmode \,$MJy\,sr\mo$\,mK$_{\rm CMB}\mo\else \,MJy\,sr\mo\,mK$_{\rm CMB}\mo$\fi}
\def\microns{\ifmmode \,\mu$m$\else \,$\mu$m\fi}

\def\muK{\ifmmode \,\mu$K$\else \,$\mu$\hbox{K}\fi}
\def\microK{\ifmmode \,\mu$K$\else \,$\mu$\hbox{K}\fi}
\def\muW{\ifmmode \,\mu$W$\else \,$\mu$\hbox{W}\fi}
\def\kms{\ifmmode $\,km\,s$^{-1}\else \,km\,s$^{-1}$\fi}
\def\kmsMpc{\ifmmode $\,\kms\,Mpc\mo$\else \,\kms\,Mpc\mo\fi}
\providecommand{\sorthelp}[1]{}

\def\WMAP{\emph{WMAP}}

\def\commander{\texttt{Commander}}

\def\commandertwo{\texttt{Commander2}}
\def\commanderthree{\texttt{Commander3}}

\renewcommand{\d}[0]{\vec{d}}
\renewcommand{\b}[0]{\vec{b}}
\newcommand{\x}[0]{\vec{x}}

\newcommand{\A}[0]{\tens{A}}
\newcommand{\Y}[0]{\tens{Y}}
\newcommand{\n}[0]{\vec{n}}

\newcommand{\gray}[0]{\color{gray}}

\newcommand{\s}[0]{\vec{s}}
\renewcommand{\a}[0]{\vec{a}}
\newcommand{\m}[0]{\vec{m}}

\newcommand{\F}[0]{\tens{F}}
\newcommand{\B}[0]{\tens{B}}

\renewcommand{\L}[0]{\tens{L}}
\newcommand{\g}[0]{\vec{g}}
\newcommand{\N}[0]{\tens{N}}
\newcommand{\M}[0]{\tens{M}}

\renewcommand{\S}[0]{\tens{S}}
\renewcommand{\r}[0]{\vec{r}}

\renewcommand{\P}[0]{\tens{P}}

\newcommand{\Dbp}[0]{\Delta_{\mathrm{bp}}}

\newcommand{\BP}{\textsc{BeyondPlanck}}

\newcommand{\npipe}[0]{\texttt{NPIPE}}

\def\inv{^{-1}}

\begin{document}

\title{\BP\ III. Commander3}
%This author list corresponds to \title{Author list for L04\_CMB\_Foregrounds\_Extraction}
%Prepared by M. Lopez-Caniego (Marcos.Lopez.Caniego@sciops.esa.int), ESAC/ESA
%This version is from Thu Jul 12 18:11:48 2018 CET
%\subtitle{There are 152 co-authors in this list}
\newcommand{\nersc}[0]{1}
\newcommand{\princeton}[0]{2}
\newcommand{\helsinkiA}[0]{3}
\newcommand{\milanoA}[0]{4}
\newcommand{\triesteA}[0]{5}
\newcommand{\haverford}[0]{6}
\newcommand{\helsinkiB}[0]{7}
\newcommand{\triesteB}[0]{8}
\newcommand{\milanoB}[0]{9}
\newcommand{\milanoC}[0]{10}
\newcommand{\oslo}[0]{11}
\newcommand{\jpl}[0]{12}
\newcommand{\mpa}[0]{13}
\newcommand{\planetek}[0]{14}
\author{\small
M.~Galloway\inst{\oslo}\thanks{Corresponding author: M.~Galloway; \url{mathew.galloway@astro.uio.no}}
\and
K.~J.~Andersen\inst{\oslo}
\and
\textcolor{black}{R.~Aurlien}\inst{\oslo}
\and
\textcolor{black}{R.~Banerji}\inst{\oslo}
\and
M.~Bersanelli\inst{\milanoA, \milanoB, \milanoC}
\and
S.~Bertocco\inst{\triesteB}
\and
M.~Brilenkov\inst{\oslo}
\and
M.~Carbone\inst{\planetek}
\and
L.~P.~L.~Colombo\inst{\milanoA}
\and
H.~K.~Eriksen\inst{\oslo}
\and
\textcolor{black}{M.~K.~Foss}\inst{\oslo}
\and
C.~Franceschet\inst{\milanoA,\milanoC}
\and
\textcolor{black}{U.~Fuskeland}\inst{\oslo}
\and
S.~Galeotta\inst{\triesteB}
\and
S.~Gerakakis\inst{\planetek}
\and
E.~Gjerl{\o}w\inst{\oslo}
\and
\textcolor{black}{B.~Hensley}\inst{\princeton}
\and
\textcolor{black}{D.~Herman}\inst{\oslo}
\and
M.~Iacobellis\inst{\planetek}
\and
M.~Ieronymaki\inst{\planetek}
\and
\textcolor{black}{H.~T.~Ihle}\inst{\oslo}
\and
J.~B.~Jewell\inst{\jpl}
\and
\textcolor{black}{A.~Karakci}\inst{\oslo}
\and
E.~Keih\"{a}nen\inst{\helsinkiA, \helsinkiB}
\and
R.~Keskitalo\inst{\nersc}
\and
G.~Maggio\inst{\triesteB}
\and
D.~Maino\inst{\milanoA, \milanoB, \milanoC}
\and
M.~Maris\inst{\triesteB}
\and
S.~Paradiso\inst{\milanoA, \milanoB}
\and
B.~Partridge\inst{\haverford}
\and
M.~Reinecke\inst{\mpa}
\and
A.-S.~Suur-Uski\inst{\helsinkiA, \helsinkiB}
\and
T.~L.~Svalheim\inst{\oslo}
\and
D.~Tavagnacco\inst{\triesteB, \triesteA}
\and
H.~Thommesen\inst{\oslo}
\and
D.~J.~Watts\inst{\oslo}
\and
I.~K.~Wehus\inst{\oslo}
\and
A.~Zacchei\inst{\triesteB}
}
\institute{\small
Computational Cosmology Center, Lawrence Berkeley National Laboratory, Berkeley, California, U.S.A.\goodbreak
\and
Department of Astrophysical Sciences, Princeton University, Princeton, NJ 08544,
U.S.A.\goodbreak
\and
Department of Physics, Gustaf H\"{a}llstr\"{o}min katu 2, University of Helsinki, Helsinki, Finland\goodbreak
\and
Dipartimento di Fisica, Universit\`{a} degli Studi di Milano, Via Celoria, 16, Milano, Italy\goodbreak
\and
Dipartimento di Fisica, Universit\`{a} degli Studi di Trieste, via A. Valerio 2, Trieste, Italy\goodbreak
\and
Haverford College Astronomy Department, 370 Lancaster Avenue,
Haverford, Pennsylvania, U.S.A.\goodbreak
\and
Helsinki Institute of Physics, Gustaf H\"{a}llstr\"{o}min katu 2, University of Helsinki, Helsinki, Finland\goodbreak
\and
INAF - Osservatorio Astronomico di Trieste, Via G.B. Tiepolo 11, Trieste, Italy\goodbreak
\and
INAF-IASF Milano, Via E. Bassini 15, Milano, Italy\goodbreak
\and
INFN, Sezione di Milano, Via Celoria 16, Milano, Italy\goodbreak
\and
Institute of Theoretical Astrophysics, University of Oslo, Blindern, Oslo, Norway\goodbreak
\and
Jet Propulsion Laboratory, California Institute of Technology, 4800 Oak Grove Drive, Pasadena, California, U.S.A.\goodbreak
\and
Max-Planck-Institut f\"{u}r Astrophysik, Karl-Schwarzschild-Str. 1, 85741 Garching, Germany\goodbreak
\and
Planetek Hellas, Leoforos Kifisias 44, Marousi 151 25, Greece\goodbreak
}

\authorrunning{BeyondPlanck Collaboration}
\titlerunning{\commander3}

\abstract{We describe the computational infrastructure for end-to-end
  Bayesian CMB analysis implemented by the \BP\ collaboration. This
  code is called \commanderthree, and provides a statistically
  consistent framework for global analysis of CMB and microwave
  observations, and may be useful for a wide range of legacy, current,
  and future experiments. The paper has three main goals. Firstly, we
  provide a high-level overview of the existing code base, aiming to
  guide readers who wish to extend and adapt the code according to
  their own needs, or to reimplement it from scratch in a different
  programming language. Secondly, we discuss some critical computational
  challenges that arise within any global CMB analysis framework, for
  instance in-memory compression of time-ordered data, FFT
  optimization, and parallelization and load-balancing. Thirdly, we
  quantify the CPU and RAM requirements for the current \BP\ analysis,
  and find that a total of 1.5\,TB of RAM is required for efficient
  analysis, and the total cost of a full Gibbs sample is 170\,CPU-hrs,
  including both low-level processing and high-level component
  separation, which is well within the capabilities of current
  low-cost computing facilities. The existing code base is made
  publicly available under a GNU General Public Library (GPL)
  license.}


\maketitle

\tableofcontents

\section{Introduction}
\label{sec:introduction}

The aim of the \BP\ project \citep{BP01} is to build an end-to-end
Bayesian CMB analysis pipeline that constrains high-level products,
such as astrophysical component maps and cosmological parameters,
directly from raw uncalibrated time-ordered data, and apply this to
the \Planck\ LFI data. This pipeline builds on the experience gained
throughout the official \Planck\ analysis period, and seeks to
translate this experience into reusable and computationally efficient
computer code that can be used for end-to-end analysis of legacy,
current, and future data sets. As a concrete, and particularly
important example, it will serve as the computational framework for
the \textsc{Cosmoglobe}\footnote{\url{https://cosmoglobe.uio.no}} effort,
which aims to establish a consistent global model of the radio,
microwave, and sub-millimeter sky through joint analysis of all
available state-of-the-art data sets. This paper gives an overview of
the \BP\ computational infrastructure, and it details several
computational techniques that allow the full exploration of the global
posterior distribution in a timely manner.

Since the beginning of precision CMB cosmology, algorithm development
has been a main focus of the community. For instance, during the early
days of CMB analysis, many different approaches to mapmaking were
explored. Projects such as \emph{COBE} \citep{cobe1,cobe2}, MAX
\citep{max}, Saskatoon \citep{saskatoon} and Tenerife \citep{tenerife}
used a wide variety of techniques, and optimality was not
guaranteed. Soon, however, the community converged on Wiener filtering
as the preferred technique \citep{tegmark}, which also allowed for the
combination of multiple datasets into a single map
\citep{CobeCombined}.

By the time \WMAP\ and its contemporaries were observing, the field
had matured to the point that common tools were used between
experiments. HEALPix\footnote{\url{http://healpix.jpl.nasa.gov}} \citep{healpix} became a \emph{de facto} standard
for pixelizing the sky, and many experiments began to use Conjugate
Gradient (CG) mapmakers on a regular basis \cite[e.g.,][]{wmap}. These
ideas were refined during the analysis of
\Planck\ \citep{planck2013-p01,planck2014-a01,planck2016-l01}, and
since then those efforts have dominated the field. Many mapmaking
tools that were developed for \Planck\ have a strong
influence on \BP, including the MADAM destriper \citep{madam}, the
LevelS simulation codebase \citep{LevelS}, the \texttt{libsharp} spherical
harmonic transform library \citep{libsharp}, and the \Planck\ DR4
(often called \npipe) analysis
pipeline \cite{npipe}.

In parallel to these mapmaking developments, algorithms for component
separation have also been gradually refined. Several different classes of
methods have been explored and applied to a variety of experiments,
including Internal Linear Combination (ILC) methods such as \WMAP\ ILC
\citep{WMAPilc1,WMAPilc2} and NILC \citep{NILC}; template-based
approaches such as SEVEM \citep{SEVEM}; spectral matching techniques
such as SMICA \citep{SMICA}; blind techniques, such as GMCA
\citep{GMCA} or FastICA \citep{fastICA}; and parametric Bayesian
modelling techniques, such as
\commander\ \citep{eriksen:2004,commander,seljebotn:2019}. This
flowering of options provided a range of complementary approaches that
each gave new insights into the underlying statistical problem.

Simultaneously, there has been an immense development in computer
hardware, increasing the amount of available CPU cycles and RAM by
many orders of magnitude. As an example, the \emph{COBE} analysis was
in 1990 initially performed on VAXstation 3200 computers
\citep{cobeprocessing}, which boasted 64 KB of RAM and a single
11\,MHz processor. For comparison, the \Planck\ FFP8 simulations
\citep{planck2014-a14} were in 2013 produced on a distributed Cray
XC30 system with 133\,824 cores, each with a clock speed of 2.4\,GHz
and 2\,GB of RAM, at a total computational cost of 25\, million CPU
hours. While the evolution in CPU clock speed has largely stagnated
during the last decade, the cost of RAM continues to decrease, and
this has been exploited to improve the memory management efficiency in
the current analysis: \BP\ represents the first CMB analysis pipeline
for which the full \Planck\ LFI time-ordered data set may be stored in
RAM on a single compute node, effectively alleviating the need for
expensive disk and network communication operations during the
analysis. As a result, the full computational cost of the current
\BP\ analysis is only 300\,000 CPU hours, and, indeed, it is not
entirely inconceivable that this analysis could be run on a laptop in
the not too distant future.

The \BP\ pipeline is a natural evolution and fusion of a wide range of
these developments into a single integrated codebase. There are
relatively few algorithmically novel features in this pipeline as
such, but the \BP\ pipeline primarily combines industry standard methods
into one single framework. The computer code that realizes this is
called \commanderthree, which is a direct generalization of
\commandertwo. Whereas previous \commander\ versions focused primarily
on high-level component separation and CMB power spectrum estimation
applications \citep{eriksen:2004,eriksen2008,seljebotn:2019},
\commanderthree\ also accounts for low-level time-ordered data
processing and mapmaking. This integrated approach yields a level of
performance and error propagation fidelity that we believe will be
difficult to replicate with distributed methods that require
intermediate human interaction. This paper describes the code
implementation that is used to produce the results detailed in the
\BP\ paper suite, and make the results of that development available
to the community. 

The \BP\ pipeline, documentation and data are all available through
the project webpage.\footnote{\url{beyondplanck.science}} In addition
to the actual \commanderthree\ source
code,\footnote{\url{https://github.com/Cosmoglobe/Commander}} several
utilties are also provided that facilitate easy use of the codes by
others in the community, for instance for downloading data and
compiling and running the codes. Documentation is also
available.\footnote{\url{docs.beyondplanck.science}} The entire
project is available under the GNU General Public License (GPL).  For
further details regarding these aspects, see \citet{BP05}.

\section{Bayesian CMB analysis, Gibbs sampling, and code design}
\label{sec:gibbs}

The main goals of the current paper are, firstly, to provide sufficient
intuition regarding the \commanderthree\ source code to allow external
users to navigate and modify it themselves, and, secondly, to present
various computational techniques that are used to optimize the
calculations. To set the context of this work, we begin by briefly
summarizing the main computational ideas behind this approach.

\subsection{The \BP\ data model and Gibbs chain}
\label{sec:bp}

As described by \citet{BP01}, \commanderthree\ is the first end-to-end
Bayesian analysis framework for CMB experiments, implementing full
Monte Carlo Markov Chain (MCMC) exploration of a global posterior
distribution. The most important component in this framework is an
explicit parametric model. The current \BP\ project primarily focuses
on the \Planck\ LFI measurements
\citep{planck2016-l01,planck2016-l02}, and for this data set we find
that the following model represents a good description of the
available measurements \citep{BP01},
\begin{equation}
  \begin{split}
    d_{j,t} = g_{j,t}&\P_{tp,j}\left[ \B^{\mathrm{symm}}_{pp',j}\sum_{c}
      \M_{cj}(\beta_{p'}, \Dbp^{j})a^c_{p'}  + \B^{\mathrm{asymm}}_{pp',j}\left(s^{\mathrm{orb}}_{j,t}  
      + s^{\mathrm{fsl}}_{j,t}\right)\right] + \\
%    + s^{\mathrm{fsl}}_{j,t} + s^{\mathrm{mono}}_{j}\right] + \\
    + &s^{\mathrm{1\,Hz}}_{j,t} + n^{\mathrm{corr}}_{j,t} + n^{\mathrm{w}}_{j,t}.
  \end{split}
  \label{eq:todmodel}
\end{equation}
Here $j$ represents a radiometer label, $t$ indicates a single
time sample, $p$ denotes a single pixel on the sky, and $c$ represents
one single astrophysical signal component. Further,
\begin{itemize}
\item $d_{j,t}$ denotes the measured data value in units of V, the calibrated timestream as output from the instrument;

\item $g_{j,t}$ denotes the instrumental gain in units of V\,K$_{\mathrm{CMB}}^{-1}$, and the specific details are discussed by \citet{BP07};

\item $\P_{tp,j}$ is a $N_{\mathrm{TOD}}\times 3N_{\mathrm{pix}}$
  pointing matrix, which in practice is stored as a compressed
  pointing and polarization angle timestream per detector (see Sect.~\ref{sec:compression}). Pointing uncertainties are currently not propagated for LFI, but a sampling
  step could be added here in future projects to include the effects
  of, for example, pointing jitter or half-wave plate uncertainties;
  
\item $\B_{pp',j}$ denotes the beam convolution term, where the asymmetric part is only calculated for the orbital dipole and sidelobe terms. This is also not sampled in the Gibbs chain currently but could be if it was possible to construct a parameterized beam model;

\item $\M_{cj}(\beta_{p}, \Dbp)$ denotes element $(c,j)$ of an
  $N_{\mathrm{det}}\times N_{\mathrm{comp}}$ mixing matrix, describing the amplitude of component $c$ as
  seen by radiometer $j$ relative to some reference frequency $\nu_0$
  when assuming some set of bandpass correction parameters $\Dbp$. Sampling this mixing matrix and the amplitude parameters is what is traditionally regarded as component separation, and is detailed by \citet{BP13} and \citet{BP14}. The sampling of the bandpass correction terms is described in \citet{BP09};
  
\item $a^c_{p}$ is the amplitude of component $c$ in pixel $p$,
  measured at the same reference frequency as the mixing matrix $\M$. The estimation of these amplitude parameters is also described by \citet{BP13} and \citet{BP14};
  
\item $s^{\mathrm{orb}}_{j,t}$ is the orbital CMB dipole signal, including relativistic quadrupole corrections. Estimation of the orbital dipole is described by \citet{BP08};
  
\item $s^{\mathrm{fsl}}_{j,t}$ denotes the contribution from far sidelobes, which is also described in \citet{BP08};

\item $s^{\mathrm{1hz}}_{j,t}$ accounts for a 1\,Hz electronic spike
  signal in the LFI detectors \citep{BP01};
  
\item $n^{\mathrm{corr}}_{j,t}$ denotes correlated instrumental noise, as is described by \citet{BP06}; and
  
\item $n^{\mathrm{w}}_{j,t}$ is uncorrelated (white) instrumental noise, which is not sampled and is simply left to average down in the maps.
\end{itemize}

Let us denote the set of all free parameters in Eq.~\eqref{eq:todmodel}
by $\omega$, such that $\omega\equiv\{g,\Dbp,\n_{\mathrm{corr}}, \a_i,
\beta_i,\ldots\}$. In that case, Bayes' theorem states that the
posterior distribution may be written in the form
\begin{equation}
  P(\omega\mid \d) = \frac{P(\d\mid \omega)P(\omega)}{P(\d)} \propto
  \mathcal{L}(\omega)P(\omega),
  \label{eq:jointpost}
\end{equation}
where $P(\d\mid \omega)\equiv\mathcal{L}(\omega)$ is called the
likelihood; $P(\omega)$ is called the prior; and $P(\d)$ is a
normalization factor.

Clearly, $P(\d\mid \omega)\equiv\mathcal{L}(\omega)$ is a complicated
multivariate probability distribution that accounts for millions of
correlated parameters, and its exploration therefore represents a
significant computational challenge. To efficiently explore this
distribution, \commanderthree\ relies heavily on Gibbs sampling
theory, which states that samples from a joint distribution may be
produced by iteratively drawing samples from all corresponding
conditional distributions. For \BP\, this translates into the
following Gibbs chain:
\begin{alignat}{10}
\label{eq:gibbsstart}
\g &\,\leftarrow P(\g&\,\mid &\,\d,&\, & &\,\xi_n, &\,\Dbp, &\,\a, &\,\beta, &\,C_{\ell})\\
\n_{\mathrm{corr}} &\,\leftarrow P(\n_{\mathrm{corr}}&\,\mid &\,\d, &\,\g, &\,&\,\xi_n,
&\,\Dbp, &\,\a, &\,\beta, &\,C_{\ell})\\
\xi_n &\,\leftarrow P(\xi_n&\,\mid &\,\d, &\,\g, &\,\n_{\mathrm{corr}}, &\,
&\,\Dbp, &\,\a, &\,\beta, &\,C_{\ell})\\
%\s^{\mathrm{mono}} &\,\leftarrow P(\s^{\mathrm{mono}}&\,\mid &\,\d, &\,\g, &\,\n_{\mathrm{corr}}, &\,\xi_n,
%&\,&\,\Dbp, &\,\a, &\,\beta, &\,C_{\ell})\\
\Dbp &\,\leftarrow P(\Dbp&\,\mid &\,\d, &\,\g, &\,\n_{\mathrm{corr}}, &\,\xi_n,
&\,&\,\a, &\,\beta, &\,C_{\ell})\\
\beta &\,\leftarrow P(\beta&\,\mid &\,\d, &\,\g, &\,\n_{\mathrm{corr}}, &\,\xi_n,
&\,\Dbp, & &\,&\,C_{\ell})\\
\a &\,\leftarrow P(\a&\,\mid &\,\d, &\,\g, &\,\n_{\mathrm{corr}}, &\,\xi_n,
&\,\Dbp, &\,&\,\beta, &\,C_{\ell})\\
C_{\ell} &\,\leftarrow P(C_{\ell}&\,\mid &\,\d, &\,\g, &\,\n_{\mathrm{corr}}, &\,\xi_n,
&\,\Dbp, &\,\a, &\,\beta&\,\phantom{C_{\ell}}),
\label{eq:gibbschain}
\end{alignat}
where the symbol ``$\leftarrow$'' means setting the variable on the
left-hand side equal to a sample from the distribution on the
right-hand side of the equation. Thus, each free parameter in
Eq.~\eqref{eq:todmodel} corresponds to one sampling step in this Gibbs
loop.

A single iteration through the main loop produces one full joint
sample, which is defined as one realization of each free
parameter in the data model. An ensemble of these samples is obtained
by running the loop for many iterations, and this ensemble can then be
used to estimate various summary statistics, such as the posterior
mean of each parameter or its standard deviation. With a sufficiently
large number of samples, one will eventually map out the entirety of
the $N$-dimensional posterior distribution, allowing exploration of
parameter correlations and other interesting effects.

\subsection{\commanderthree\ and object-oriented programming}

\newcommand{\find}{\quad}
\newcommand{\ind}{\quad \;}
\begin{lstfloat}[t]
  {\scriptsize
    \begin{tcolorbox}
\begin{lstlisting}
i)   Read parameter file 
ii)  Initialize data sets; store in global array
       called data
iii) Initialize model components; store in global 
       linked list called compList
iv)  Initialize stochastic parameters 

for i = 1, N_gibbs do
  1) Process TOD into frequency maps
     a) Sample gain
     b) Sample correlated noise
     c) Clean and calibrate TOD
     d) Sample bandpass corrections
     e) Bin TOD into maps
  2) Sample astrophysical amplitude parameters
  3) Sample angular power spectra
  4) Output current parameter state to disk
  5) Sample astrophysical spectral parameters
  6) Sample global instrument parameters for non-TOD
       data sets, including calibration, bandpass
       corrections
\end{lstlisting}
    \end{tcolorbox}
}
\caption{Schematic overview of \commanderthree\ execution.}\label{code:gibbs}
\end{lstfloat}

\commanderthree\ represents a translation of the Gibbs chain shown in
Eqs.~\eqref{eq:gibbsstart}-\eqref{eq:gibbschain} into computer code. This is made more concrete
in Listing~\ref{code:gibbs} in terms of high-level pseudocode. A
detailed breakdown is provided in Sect.~\ref{sec:commander3}, and here
we only make a few preliminary observations. First, we note that Gibbs
sampling naturally lends itself to object oriented programming due to
its modular nature. Each component in the Gibbs chain can typically be
compartmentalized in terms of a class object, and this greatly
alleviates code complexity and increases modularity.

\commanderthree\ is designed with this philosophy in mind, while at
the same time optimizing efficiency through the use of some key global
variables. The two most important global objects of this type are
called \texttt{data} and \texttt{compList}. The first class provides
convenient access to all data sets included in the analysis (e.g.,
\Planck\ 30\,GHz or \WMAP\ K-band data). Classes are provided both for
high- and low-level data objects. An example of the former is
\texttt{comm\_tod\_noise\_mod} which provides routines for sampling
the correlated noise parameters of a given dataset, while
\texttt{comm\_tod\_orbdipole\_mod} calculates the orbital dipole
estimate. An
example of the latter is \texttt{comm\_map\_mod}, which corresponds to
a HEALPix map object, stored either in pixel or harmonic space. The
same class also provides map-based manipulation class
functions, for instance spherical harmonic transforms (SHT) routines
or smoothing operators.

The second main variable, \texttt{compList}, is a linked list of all
model component objects, describing for instance CMB or
synchrotron emission. Again, each class contains both the infrastructure and
variables needed to define the object in question, and the required
sampling routines for the respective free variables. For instance,
\texttt{comm\_comp\_mod} represents a generic astrophysical sky
component, while a specific subclass such as
\texttt{comm\_freefree\_comp\_mod} represents free-free emission
specifically.  Another example is \texttt{comm\_Cl\_mod}, which
defines angular power spectra, $C_\ell$, and provides the required
sampling routines for this.

A third type of \commanderthree\ modules is more diverse, and provides
general infrastructure support. Examples include wrappers for
underlying libraries or functionality like \texttt{comm\_hdf\_mod}
(used for IO operations), \texttt{comm\_huffman\_mod} (used for
in-memory data compression), or \texttt{sharp.f90} (used for spherical
harmonics transforms). Other modules, like \texttt{comm\_utils} or
\texttt{math\_tools}, provide general utility functions, for instance
for reading simple data files or inverting matrices. Ultimately this
category of classes is a concession to the fact that not all
functionality need be encapsulated in a truly object oriented way. 

Returning to Listing~\ref{code:gibbs}, we see that \commanderthree\ may
be summarized in terms of two main phases, namely initialization and
execution. The goal of the initialization phase is simply to set up
the \texttt{data} and \texttt{compList} objects, while the execution
phase essentially amounts to repeated updates of the stochastic object
variables that are stored within each of these objects. Finally, the
current state of those variables are stored to disk at regular
intervals, resulting in a Markov chain of parameter states. 

\subsection{Memory management and parallelization}
\label{sec:memory}

Essentially all of the above considerations apply equally well to
\commandertwo\ \citep{seljebotn:2019} as to \commanderthree, as the
only fundamentally new component in the current analysis is additional
support for time-ordered data processing. However, this extension is
indeed nontrivial, as it has major implications in terms of
computational efficiency and memory management. In particular, while
traditional Bayesian component separation (as implemented in
\commandertwo) is limited by the efficiency of performing spherical
harmonics transforms, TOD processing is strongly dominated by memory
bus efficiency, i.e., by the cost of shipping data from RAM to the
CPU. These two problems clearly prefer different parallelization and
load-balancing paradigms, and combining the two within a single
framework represents a notable challenge.

As a temporary solution to this problem, the current
\BP\ analysis \citep{BP01} is run on a small-sized cluster hosted by
the University of Oslo that consists of two compute nodes with each
128 AMD EPYC 7H12 2.6\,GHz cores and 2\,TB of RAM. This amount of RAM
allows storage of the full TOD on each node, and each node runs a
completely independent Gibbs chain. As a result, any communication
overhead is entirely eliminated, resulting in high overall efficiency.

\begin{figure*}[t]
  \center
  \includegraphics[angle=270,width=\linewidth]{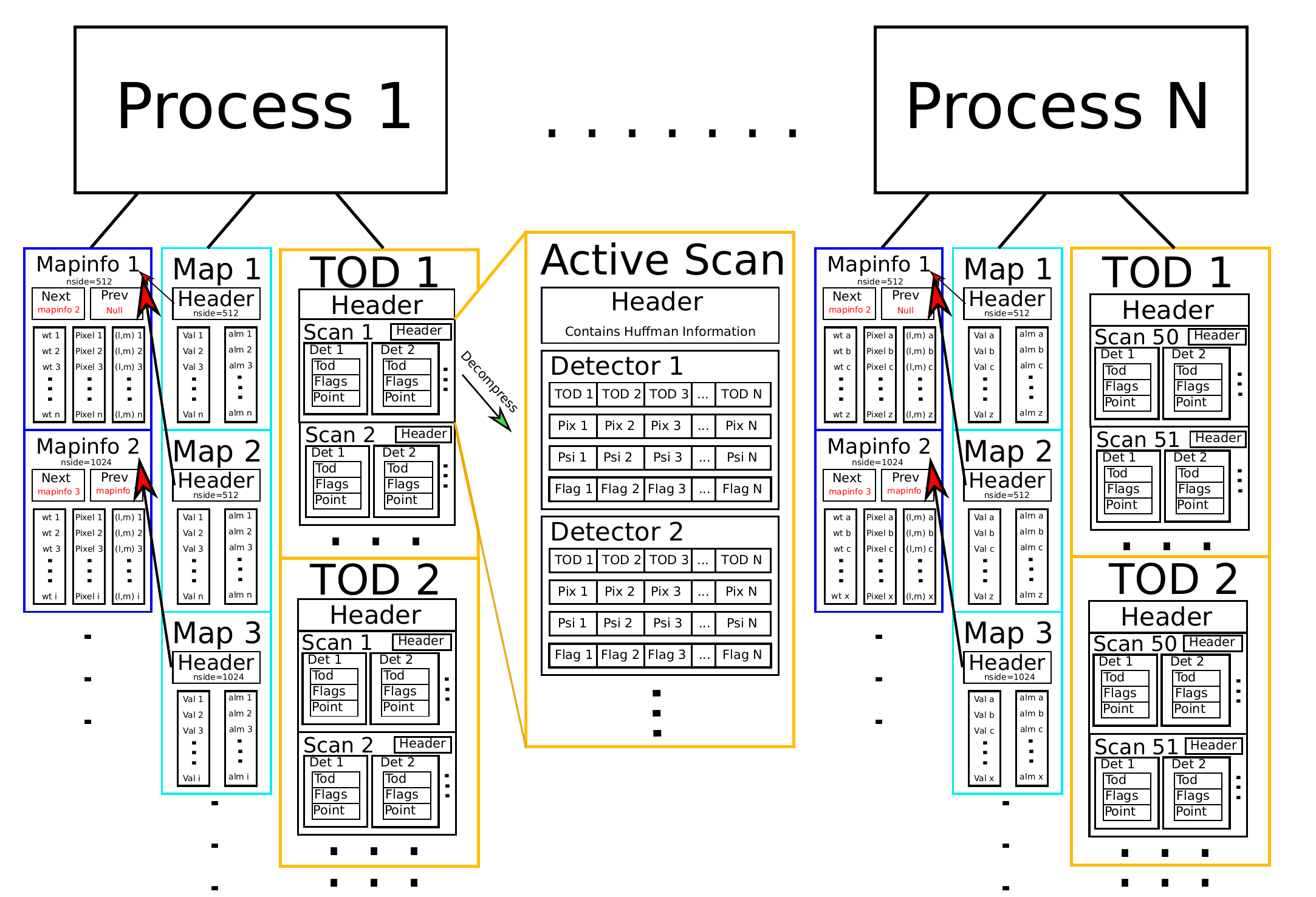}
  \caption{\commanderthree\ memory layout for map and TOD objects. The
    top \texttt{process} box represents a single computing core. 
  }\label{fig:memory}
\end{figure*}

However, \commanderthree\ is parallelized using the Message Passing
Interface (MPI), adopting a ``crowd computing, node-only''
parallelization paradigm, in which all cores participate equally in
most computational operations. As such, the code can technically run
on massively distributed systems with limited RAM per node, and this mode
of operation will clearly be needed for applications to large data sets,
such as those produced by ground-based experiments (e.g., Simons
Observatory or CMB-S4; \citealp{simons,s4}), which will require hundreds of TB of RAM and
hundreds of thousands of cores. However, while the existing code may
run in this mode, it will clearly not be computationally efficient,
because of the flat parallelization paradigm: The actual run time will
be massively dominated by network communication in spherical harmonics
transforms, to the point that the code is useless. As such, a
dedicated rewrite of the underlying parallelization infrastructure is
certainly required for efficient end-to-end Bayesian analysis of large
volume data sets; the current \commanderthree\ implementation is
rather tuned for TB-sized data sets, such C-BASS \citep{cbass}, \Planck, SPIDER \citep{spider},
\WMAP\ \citep{wmap} --- and in a few years, possibly even LiteBIRD \citep{LiteBIRD}.

Consequently, the current code is typically run with
$\mathcal{O}(100)$ cores per chain, which is determined by the
requirement of achieving good spherical harmonics efficiency for all
data maps involved in the analysis. In order to increase the overall
concurrency, it is typically computationally advantageous to run more
independent Markov chains in parallel, rather than adding more cores
to each chain. Efficient in-chain parallelization is achieved through
data distribution across cores, such that each process is given only a
subset of each dataset to operate on locally, and results are shared
between processes only when absolutely necessary.

\subsection{Memory layout}
\label{sec:mem_layout}

As already mentioned, the single most important bottleneck for this
analysis is shipping data from RAM to the CPU, and an efficient memory
layout is therefore essential to maintain high throughput. The layout
adopted for \commanderthree\ is schematically illustrated in
Fig.~\ref{fig:memory}. In particular, there are two main types of data
that need to be distributed, namely maps and TOD. During
initialization, each process is assigned a segment of each map (both
in pixel and harmonic space) and a set of time chunks of each TOD
object; for \Planck\ LFI, these are for convenience distributed
according to pointing periods.

Map objects (shown as light blue boxes in Fig.~\ref{fig:memory}) are
used to represent astrophysical components, spatially varying
parameters, beam functions and many other things, and are distributed
according to the \texttt{libsharp} parallelization scheme
\citep{libsharp}. This choice is based on the fact that
\texttt{libsharp} is the most efficient spherical harmonics transform
library available in the CMB field today, and optimizing this
operation is essential. Each map object can simultaneously be
expressed as pixels or a set of $a_{\ell,m}$ components in harmonic
space, and \texttt{libsharp} uses fast SHTs to convert between these
two representations. Each core is given a subset of pixels and
$a_{\ell,m}$ coefficients, and for two maps with the same HEALPix
resolution $\N_{\mathrm{side}}$, any given core will receive exactly
the same pixels of each map, which helps minimize the amount of
overhead for cross-frequency local operations. The header for each map
(which includes information such as $\N_{\mathrm{side}}$,
$\ell_{\mathrm{max}}$, $n_{\mathrm{map}}$,\footnote{The
  $n_{\mathrm{map}}$ parameter defines the number of columns in a
  given map, which typically corresponds to the number of Stokes
  parameters.} pixel and $a_{\ell m}$ lists for current core etc.) is
stored as a pointer to an object called \texttt{mapinfo} (dark blue in
Fig.~\ref{fig:memory}), that itself is only stored once per unique
combination of $\{\N_{\mathrm{side}}$,$\ell_{\mathrm{max}}$,
$n_{\mathrm{map}}\}$ to save memory.

A single TOD object (shown as yellow boxes in Fig.~\ref{fig:memory})
represents all time ordered data (TOD) from one set of detectors at a common
frequency or, equivalently, all the data one would want to combine
into a single frequency map. These objects are generally very large,
to the point where it is barely feasible to hold a single copy in
memory. Therefore, the TODs are divided into discrete time chunks of a
reasonable length, and distributed across cores. To minimize the
memory footprint, all ancillary TOD objects (for instance, flags and
pointing) are stored in memory in compressed format, as described in
Sect.~\ref{sec:compression}, and must be decompressed before any
timestream operations can be performed. Thus, each chunk is processed
sequentially, with the first step being decompression into standard
time ordered arrays with common indexing. Those local data objects are
then utilized and cleaned up before processing the next chunks. All
inputs required for global TOD operations (such as gain sampling or
map binning; see Step~2 in Listing~\ref{code:gibbs}), are coadded on the
fly during this iteration over chunks, and synchronization across cores
is done only once after the full iteration.

Finally, to further minimize the memory footprint during the TOD
binning phase (during which each core in principle needs access to
pixels across the full sky), TODs are distributed according to their
local sky coverage. For \Planck, this implies that any single core
processes pointing periods for which the satellite spin axis are
reasonably well aligned, and the total sky coverage per core is typically
10\,\% or less. This also minimizes network communication overhead
during the synchronization stage.

\begin{lstfloat}[t]
  {\scriptsize
    \begin{tcolorbox}
\begin{lstlisting}
**************************************************
*           Commander parameter file             *
**************************************************
@DEFAULT LFI_tod.defaults

OPERATION            = sample  # {sample,optimize}

##################################################
#            Algorithm specification             #
##################################################

# Monte Carlo options
NUMCHAIN         = 1     # Number of independent chains
NUM_GIBBS_ITER   = 1500  # Length of each Markov chain
INIT_CHAIN01     = /path/to/chain/chain_init_v1.h5:1

SAMPLE_SIGNAL_AMPLITUDES      = .true.
SAMPLE_SPECTRAL_INDICES       = .true.
ENABLE_TOD_ANALYSIS           = .true.

##################################################
#                Output options                  #
##################################################

OUTPUT_DIRECTORY              = chains_BP10

##################################################
#                Data sets                       #
##################################################

DATA_DIRECTORY                 = /path/to/workdir/data
NUMBAND                        = 2

INCLUDE_BAND001                = .true.   # 30 GHz
INCLUDE_BAND002                = .true.   # 44 GHz

# 30 GHz parameters
@START 001
@DEFAULT bands/LFI/LFI_030_TOD.defaults
BAND_MAPFILE&&&                = map_030_BP10.1_v1.fits
BAND_NOISEFILE&&&              = rms_030_BP10.1_v1.fits
BAND_TOD_START_SCANID&&&       = 3
BAND_TOD_END_SCANID&&&         = 44072
@END 001

# 44 GHz parameters
@START 002
@DEFAULT bands/LFI/LFI_044_TOD.defaults
BAND_MAPFILE&&&                = map_044_BP10.1_v1.fits
BAND_NOISEFILE&&&              = rms_044_BP10.1_v1.fits
@END 002

##################################################
#              Model parameters                  #
##################################################

NUM_SIGNAL_COMPONENTS     = 2
INCLUDE_COMP01            = .true.  # CMB 
INCLUDE_COMP02            = .true.  # Synchrotron

# CMB
@START 01
@DEFAULT components/cmb/cmb_LFI.defaults
COMP_INPUT_AMP_MAP&&      = cmb_amp_BP8.1_v1.fits
COMP_MONOPOLE_PRIOR&&     = monopole-dipole:mask.fits
@END 01

# Synchrotron component
@START 02
@DEFAULT components/synch/synch_LFI.defaults
COMP_INPUT_AMP_MAP&&          = synch_amp_BP8.1_v1.fits
COMP_INPUT_BETA_MAP&&         = synch_beta_BP8.1_v1.fits
COMP_PRIOR_GAUSS_BETA_MEAN&&  = -3.3
@END 02
\end{lstlisting}
    \end{tcolorbox}
}
\caption{Prototype \commander\ parameter file.}\label{listing:param}
\end{lstfloat}

\section{The \commanderthree\ software}
\label{sec:commander3}

In this section, we give a walkthrough of the
\commanderthree\ software package, organized roughly according to the
order in which a new user will experience the code. That is, we start
with the code base, installation procedure, and documentation, before
describing the \commander\ parameter file. Then we consider the actual
code, and describe the main modules. 

\subsection{Code base, documentation, installation, and execution}

The \commander\ code base, installation procedure, and documentation
is described by \citet{BP05}, with a particular emphasis on
reproducability. In short, the code is made publicly available on
GitHub\footnote{\url{https://github.com/Cosmoglobe/Commander.git}}
under a GNU General Public Library (GPL) license, and the
documentation\footnote{\url{https://cosmoglobe.github.io/Commander/}}
is also hosted at the same site.

At present, only Linux- and MPI-based systems are supported, and the
actual installation procedure is CMake-based, and may in an ideal
case be as simple as executing the following command line commands:
{\small
\begin{verbatim}
> git clone https://github.com/Cosmoglobe/Commander.git 
> mkdir Commander/build && cd Commander/build
> cmake -DCMAKE_INSTALL_PREFIX=$HOME/local \
        -DCMAKE_C_COMPILER=icc \
        -DCMAKE_CXX_COMPILER=icpc \
        -DCMAKE_Fortran_COMPILER=ifort \
        -DMPI_C_COMPILER=mpiicc \
        -DMPI_CXX_COMPILER=mpiicpc \
        -DMPI_Fortran_COMPILER=mpiifort \
        ..
> cmake --build . --target install -j 8
\end{verbatim}
} In this particular example, we use an Intel compiler suite, but the
code has also been tested with GNU compilers. The first command
downloads the source code; the second command creates a local
directory for the specific compiled version; the third command creates
a CMake system-specific compilation recipe (similar to Makefile) that
accounts for all dependent libraries, such as
HEALPix,
FFTW,\footnote{\url{https://fftw.org}} \texttt{libsharp} etc.; and the
fourth command actually downloads and compiles all required libraries
and executables. In practice, problems typically do emerge on any new
system, and we refer the interested (or potentially frustrated) reader
to the full documentation for further information.

Once the code is successfully compiled, it is run through the system
MPI environment, for instance
\begin{verbatim}
mpirun -n {ncore} path/to/commander param.txt
\end{verbatim}
Specific MPI runtime environment parameters must be tuned according to
the local system.

\subsection{The \commander\ parameter file}

After successfully compiling and running the code, the next step in
the process encountered by most users is to understand the
\commander\ parameter file. This is a simple human readable and
editable ASCII file with one parameter per line, of the form
\begin{verbatim}
PARAMETER_NAME = {value}
\end{verbatim}
The value cannot contain blank spaces (as anything following a space
in the same line is ignored, and can be used for comments) or
quotation marks, which serve a reserved internal purpose.

The \commander\ parameter file can become very long for
multi-experiment configurations, as in several thousands of lines, and
maintaining readability is essential for effective debugging and
testing purposes. For help in this respect, the \commander\ parameter
file supports four special directives that allow the construction of
nested parameter files, namely
\begin{verbatim}
  @INCLUDE {filename_with_full_path}
  @DEFAULT {filename_with_relative_path}
  @START {number}
  @END {number}
\end{verbatim}
The first two of these simply insert the full contents of the
specified parameter file at the calling location of the directive, and
the only difference between them is whether the filename specifies a
full path (as in the first case) or a path relative to a library of
default parameter files (as in the second case). Nested include
statements are allowed, and it is always the last occurrence of a given
parameter that is used. The two latter directives replace any
occurrence of multiple ampersands between \texttt{@START} and \texttt{@END} with the
specified number.

The default parameter file library is provided as part of the
\commander\ source code, and the path to this must be specified
through an environment variable called
\texttt{COMMANDER\_PARAMS\_DEFAULT}. This library contains default parameter
files for each data set (\Planck\ LFI 30\,GHz, \WMAP\ Ka-band etc.),
as well as for each supported signal component (CMB, synchrotron,
thermal dust emission etc.), and allow for simple construction of
complex analysis configurations through the use of well-defined
parameter files per data set and component. These also serve as useful
templates when adding new experiments or components.

Parameters may also be submitted with double dashes on the command
line at runtime (e.g., \texttt{-{}-BASE\_SEED=4842}), to support
convenient scripting. Any parameter submitted through the command line
takes preference over those defined in the parameter file.

The entire parameter file is parsed and validated as the very first
step of the code execution, and stored in an internal data structure
called \texttt{comm\_params} for later usage. This is done to catch
user errors early in the process, and speed up debugging and
testing. The internal parameter data structure is also automatically
written to the output directory for reproducibility purposes.

An example of a top-level \commander\ parameter file with two
frequency maps (\Planck\ LFI 30 and 44\,GHz) and two astrophysical
components (CMB and synchrotron emission) is shown in
Listing~\ref{listing:param}. A full description of all supported
parameters is provided in the online documentation referenced
above. We do note, however, that the quick pace of code development
sometimes leaves the documentation out-of-date. If this happens, we
encourage the reader to submit an issue through the GibHub repository,
or simply fix it, and submit a pull request; \commander\ is an Open Source project,
and community contributions are much appreciated.

\begin{figure*}[t]
  \center
  \includegraphics[width=\linewidth]{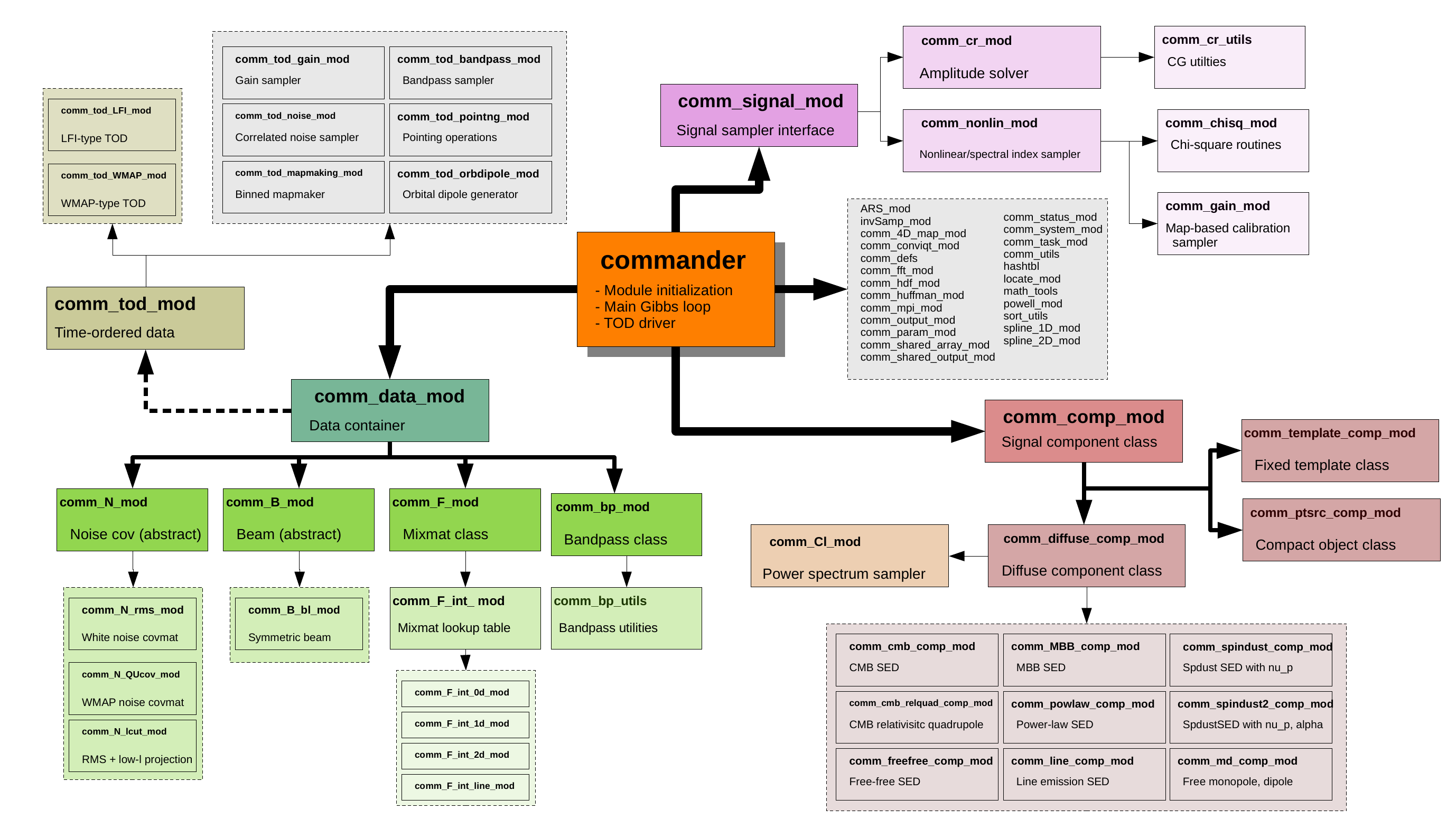}
  \caption{Overview of the \commander\ source code. The main program
    is called \texttt{commander.f90}, and is indicated by the orange
    box in the center. All other boxes represents individual modules,
    except the gray box to the right, which summarizes various utility modules.
  }\label{fig:commander}
\end{figure*}

\subsection{Source code overview}

After being able to run the code and edit the parameter file, the next
step is usually to modify the actual code according to the needs of a
specific analysis, whether it is to add support for a new
astrophysical component or a new low-level TOD processing
type. Clearly, this process may feel somewhat intimidating, given that
the current code base currently spans more than 60\,000 lines
distributed over 96 different Fortran modules. Fortunately, as already
mentioned, the code is highly modular in structure, and any given
development project can in most cases only focus on a relatively small
part of the code to achieve its goals. The goal of the current section
is to provide a ``code map'' that helps the user to navigate the code.

This map is shown in Fig.~\ref{fig:commander} in terms of main
modules. Each colored block represents one Fortran module with the
name given in bold. (The gray utility box is a special case, in which
each entry indicates a separate module.) Different colors represent
different module types, namely data objects (green), astrophysical
component objects (red), signal sampling interfaces (purple), utility
routines (gray), as well as the main program (orange). We note that
this map is not exhaustive, as new modules are added regularly.

The starting point for any new user is typically the main program
file, \texttt{commander.f90}. This file implements the overall
execution structure that was outlined in Listing~\ref{code:gibbs}, which
may be divided into module initialization and main Gibbs operations,
and spans only about 500 lines of code. From this module, one may
follow the arrows in Fig.~\ref{fig:commander} to identify any specific
submodule.

\subsubsection{Data infrastructure}

\begin{lstfloat}[t]
  {\scriptsize
    \begin{tcolorbox}
\begin{lstlisting}
  type comm_data_set
     character(len=512)           :: label
     character(len=512)           :: unit
     integer                      :: ndet
     character(len=128)           :: tod_type
     logical                      :: pol_only

     class(comm_mapinfo),pointer :: info
     class(comm_map),    pointer :: map 
     class(comm_map),    pointer :: res 
     class(comm_map),    pointer :: mask
     class(comm_map),    pointer :: procmask
     class(comm_tod),    pointer :: tod
     class(comm_N),      pointer :: N  
     class(B_ptr),       allocatable, dimension(:) :: B
     class(comm_bp_ptr), allocatable, dimension(:) :: bp
   contains
     procedure :: RJ2data
     procedure :: chisq => get_chisq
  end type comm_data_set
\end{lstlisting}
    \end{tcolorbox}
}
\caption{Prototype \commander\ data class, \texttt{comm\_data\_set}.}\label{listing:data}
\end{lstfloat}

The main data interface is defined in \texttt{comm\_data\_mod}, in
terms of a class called \texttt{comm\_data\_set}. Each object of this
type represents one frequency channel, for instance \Planck\ 30\,GHz,
\WMAP\ $Q$-band, or Haslam 408\,MHz. The main class definition is
shown in Listing~\ref{listing:data}, which is common to all data
objects. The top section defines various scalars, such as the
frequency channel label (e.g., \texttt{LFI\_030}), unit type (e.g.,
uK), number of detectors, and TOD type (if any). The next section
defines pointers to various map-level objects, including the actual
coadded frequency map, main and processing masks, and the
data-minus-model residual. All of these share the same
\texttt{mapinfo} instance, as outlined in Fig.~\ref{fig:memory},
stored in \texttt{info}.

The noise model is defined in terms of a pointer to an abstract and
generic \texttt{comm\_N} class (implemented in \texttt{comm\_N\_mod}
in Fig.~\ref{fig:commander}), which is instantiated in terms of a
specific subclass. At the moment, only three noise types are
supported, namely spatially uncorrelated white noise ($\N_{pp'} =
\sigma_p\delta_{pp'}$, implemented in \texttt{comm\_n\_rms\_mod}); a
full dense low-resolution noise covariance matrix for Stokes $Q$ and $U$,
as defined by the \WMAP\ data format, implemented in
\texttt{comm\_n\_qucov\_mat}; and white noise per pixel, but
projecting out all large-scale harmonic modes with
$\ell\le\ell_{\mathrm{cut}}$, implemented in
\texttt{comm\_n\_lcut\_mat}. Each of these modules defines routines
for multiplying a given sky map with operators like $\N^{-1}$,
$\N^{-1/2}$, and $\N^{1/2}$, but does not permit access to specific
individual elements (except for diagonal elements, which are used for
conjugate gradients preconditioning). As such, very general noise
modules may easily be defined, and external routines do not have to
care about the internal structure of the noise model. 

Next, each data object is associated with a beam operator, $\B$,
implemented in \texttt{comm\_B\_mod}. Once again, this is an abstract
class, and must be instantiated in terms of a subclass. In the
current implementation, only azimuthally symmetric beams defined by a
Legendre transform $b_{\ell}$ are supported (in
\texttt{comm\_B\_bl\_mod}), but future work may for instance aim to
implement support for asymmetric \texttt{FeBeCOP} beams \citep{mitra2010} or
time-domain total convolution \citep{Wandelt:2001,BP08}.

Bandpass integration routines are implemented through the
\texttt{comm\_bp} module, and are accessible for each data set through
the \texttt{bp} pointer. This module defines the effective bandpass,
$\tau$, per detector (if relevant) and for the full coadded frequency
channel. It provides both unit conversion factors and astrophysical
SED integration operations, adopting the notation of
\citet{planck2013-p03d,BP09}. The specific integration prescription
must be specified according to experiment; the differences between the
various cases account for instance for whether $\tau$ is defined in brightness
or flux density units, or whether any thresholds are applied to
$\tau$. In general, we choose to reimplement the conventions adopted
by each experiment individually, rather than modifying the inputs to
fit a standard convention, to stay as close as possible to the
original analyses.

Astrophysical SED bandpass integration is performed in the mixing
module class, \texttt{comm\_F\_mod}.\footnote{Mixing matrix operators
  are for historical reasons currently denoted $\F$ in \commander; it
  is likely to change to $\M$ in a future update, conforming to the more
  modern notation used in the literature.} A central step in
the \commander\ analysis is fitting spectral parameter for each
astrophysical component, and this requires repeated integration of
parametric SEDs with respect to each bandpass. To avoid performing the
full integral for every single parameter change, we precompute lookup
tables for each component and bandpass over a grid in each
parameter. For SEDs with one or two parameters, we use (bi)cubic
splines to interpolate within the grid. Separate modules are provided
for constant, one- and two-dimensional SEDs, as well as
$\delta$-function SEDs (supporting line emission components). While
this approach is computationally very efficient, it also introduces an
important limitation, in that only two- or lower-dimensional
parametric SEDs are currently supported; future work should aim to
implement arbitrary dimensional SED interpolation, for instance using
machine learning techniques \citep[e.g.,][]{fendt:2007}.

\begin{lstfloat}[t]
  {\scriptsize
    \begin{tcolorbox}
\begin{lstlisting}
! TOD class for all scans, all detectors
type, abstract :: comm_tod
   character(len=512) :: freq
   character(len=128) :: tod_type
   integer            :: nmaps  
   integer            :: ndet   
   integer            :: nscan  
   real               :: samprate
   type(comm_scan), dimension(:) :: scans
 contains
   procedure  :: read_tod
   procedure  :: process_tod
   procedure  :: decompress_tod
   procedure  :: tod_constructor
end type comm_tod
   
! ####################################################

! TOD class for single scan, all detectors
type :: comm_scan
   integer    :: ntod
   real       :: v_sun(3)
   class(comm_detscan), dimension(:) :: d
end type comm_scan

! ####################################################

! TOD class for single detector and single scan
type :: comm_detscan
   logical    :: accept
   class(comm_noise_psd), pointer   :: N_psd
   byte,               dimension(:) :: tod
   byte,               dimension(:) :: flag
   type(byte_pointer), dimension(:) :: pix
   type(byte_pointer), dimension(:) :: psi
end type comm_detscan
\end{lstlisting}
    \end{tcolorbox}
}
\caption{TOD object structure used in \commander. Note that these
  module descriptions are incomplete, and only intended to illustrate
  the structure, not the full contents. }\label{listing:tod}
\end{lstfloat}

For relevant channels, time-ordered data are stored in the abstract
\texttt{comm\_tod} class, which is illustrated in
Listing~\ref{listing:tod}. This structure has three levels. At the
highest level, \texttt{comm\_tod} describes the full TOD for all
detectors and all scans. This object defines all parameters that are
common to all detectors and scans, for instance frequency label,
sampling rate, and TOD type. It also contains an array of
\texttt{comm\_scan} objects, each of which contains the TOD of a
single scan for all detectors. This module defines all parameters that
are common to that particular scan, for instance the number of samples
in the current scan, $n_{\mathrm{tod}}$, or the satellite velocity
with respect to the Sun, $v_{\mathrm{Sun}}$. It also contains an array
of \texttt{comm\_detscan} objects, in which the actual data for a
single scan and single detector are stored. Note that the various
data, flag and pointing arrays (\texttt{tod}, \texttt{flag},
\texttt{pix}, \texttt{psi}) are stored in terms of byte objects, which
indicates that these are all Huffman compressed, as discussed in
Sect.~\ref{sec:mem_layout}. (This feature is optional, and it is
possible to store the data uncompressed.) The \texttt{comm\_tod}
object provides the necessary decompression routines.

The most important TOD routine is \texttt{process\_tod} in the
\texttt{comm\_tod} object. The main task of this routine is to produce
a sky map and its noise description given an astrophysical reference
sky. This includes both performing all relevant TOD-level sampling
steps, and solving for the actual map either through binning or
Conjugate Gradient solvers. Since each experiment in general requires
different sampling steps and mapmaking approaches, we have chosen to
implement one TOD module per experiment, for instance
\texttt{comm\_tod\_lfi\_mod}, as opposed to one super-module for all
experiments with excessive use of conditional if-tests. This both
makes the overall TOD processing code more readable, and it allows
different people to work simultaneously on different experiments with
fewer code synchronization problems.\footnote{We note that the first
  implementation of this actually did use common routines for multiple
  experiments, but this strategy was quickly abandoned due to
  complicated merging procedures.} The main costs are significant
code replication and a higher risk of code divergence during
development. Common operations are, however, put into general TOD
modules, such as \texttt{comm\_tod\_gain\_mod} and
\texttt{comm\_tod\_orbdipole\_mod}, with the goal of maximizing code
reusability.

\subsubsection{Signal model infrastructure}

The red boxes in Fig.~\ref{fig:commander} summarize modules that
define the astrophysical sky model, and the purple boxes contain
corresponding sampling algorithms. Starting with the former, we see
that three fundamentally different types of components are currently
supported, namely 1) diffuse components, 2) point source components,
and 3) template components. The first of these is by far the most
important, as it is used to describe the usual spatially varying
``full-sky'' components, such as CMB, synchrotron, thermal dust
emission etc. The main difference between the various diffuse
components are their spectral energy densities (SEDs) that defines the
signal strength as a function of frequency in units of brightness
temperature, with some set of free parameters. Examples of currently
supported SEDs are listed in the bottom right block of
Fig.~\ref{fig:commander}. We note, however, that it is very easy to
add support for a new SED type as follows. First, determine how many
free spectral parameters the new component should have; if it is less than
or equal to two, then identify an existing component with the same
number of parameters, and copy and rename the corresponding module file. Then edit the function
called \texttt{evalSED} in that routine to return the desired
parametric SED. Finally, search for all occurrences of the original
component label in the \texttt{comm\_signal\_mod} module, and add
corresponding entries for the new component. Typically, adding a new
SED type with two or fewer parameters can be done in 15 minutes; if
the component has more than two free parameters, however, new mixing
matrix interpolation and precomputation infrastructure has to be
implemented, as discussed above.

The spatial distribution of a diffuse component is defined in terms of a
spherical harmonics expansion, $s(\hat{n}) = \sum a_{\ell m} Y_{\ell
  m}(\hat{n})$, with an upper frequency cutoff, $\ell_{\mathrm{max}}$,
coupled to the SED discussed above. Optionally, the spherical
harmonics coefficients may be constrained through an angular power
spectrum, $C_{\ell}\equiv\left<|a_{\ell m}|^2\right>$, that
intuitively quantifies the smoothness of the component through the
signal covariance matrix, $\S$ \citep{BP13}. Currently supported power
spectrum modes include
\begin{itemize}
\item \texttt{binned}: $D_{\ell}\equiv C_{\ell}\ell(\ell+1)/2\pi$
is piecewise constant within user-specified bins; typically the
default choice for the CMB component \citep{BP11,BP12};
\item \texttt{power\_law}: $D_{\ell}=q(\ell/\ell_0)^{\alpha}$, where
  $q$ is an amplitude, $\ell_0$ is a pivot multipole, and $\alpha$ is
  a spectral slope; often used for astrophysical foregrounds, such as
  synchrotron or thermal dust emission \citep[e.g,][]{planck2014-a12};
\item \texttt{gauss}: $D_{\ell} = q\,\exp(-\ell(\ell+1)\sigma^2)$,
  where $\sigma$ is a user-specified standard deviation; used to
  impose a natural smoothing scale to suppress Fourier ringing;
\item \texttt{none}: no power spectrum prior is applied, $\S^{-1}=0$.
\end{itemize}
Support for integrated cosmological parameter models through CAMB
\citep{Lewis:1999bs} is ongoing. When complete, this will be added as
a new type for which $\S$ will be defined in terms of the usual
cosmological parameters ($H_0$, $\Omega_i$, $\tau$ etc.).

Point source components are defined through a user-specified catalog
of potential source locations, following \citet{planck2016-l04}. Each
source is intrinsically assumed to be a spatial $\delta$ function,
with an amplitude defined in units of flux density in
milli-Janskys. Each source location is then convolved with the local
instrumental beam shape of each channel (typically asymmetric \texttt{FeBeCOP}
beam profiles for \Planck, \citealp{mitra2010}, and azimuthally
symmetric beam profiles for other experiments), and this is adopted as
a spatial source template at the relevant frequency channel. In
addition, each source is associated with an SED, similar to the
diffuse components, allowing for extrapolation between
frequencies. Currently supported models include power-law (for radio
sources), modified blackbody (for far-infrared sources), and thermal
Sunyaev-Zeldovich SEDs. Time variability is not yet supported.

\begin{figure}[t]
  \center
  \includegraphics[width=\linewidth]{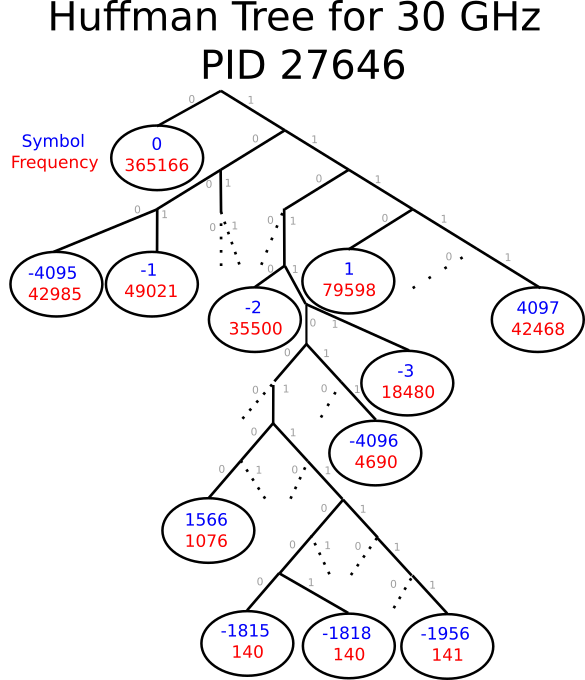}
  \caption{A truncated Huffman tree for the compression of the LFI 30 GHz channel at $\N_{\mathrm{side}}=512$ on PID 27\,646. The ovals represent the "leafs" of the tree, each of which contains a single number to be compressed. To determine the symbol that represents each number in the Huffman binary array, simply read down the tree from the top, adding a 0 for every left branch and a 1 for every right branch. The number $-2$, for example, is in this table represented by the binary code 11010. Dotted lines represent branches that were truncated for visual reasons. The full tree contains 670 unique numbers with a total array size of 861\,276 entries.
  }\label{fig:huffman}
\end{figure}

\begin{figure}[t]
  \center
  \includegraphics[width=\linewidth]{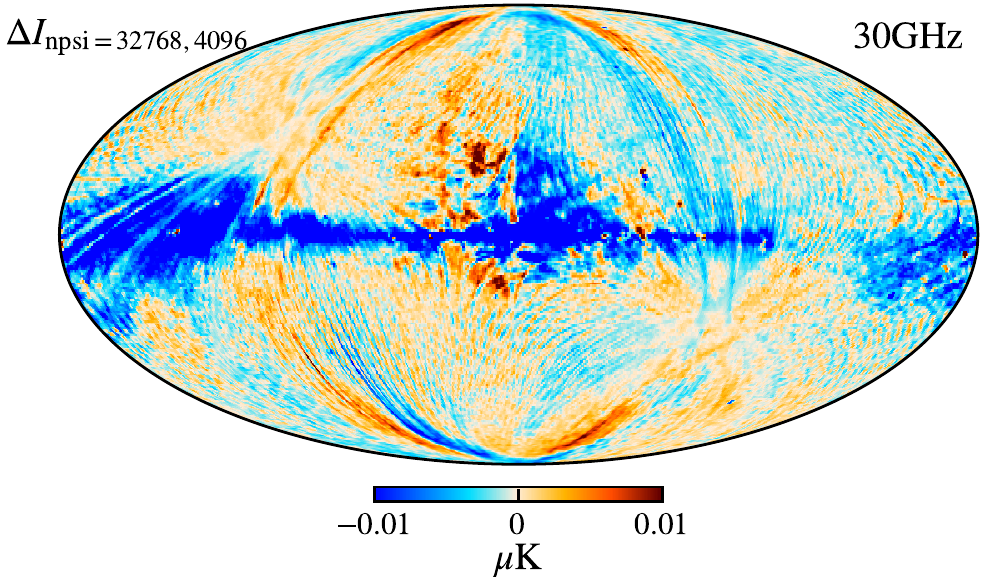}\\
  \includegraphics[width=\linewidth]{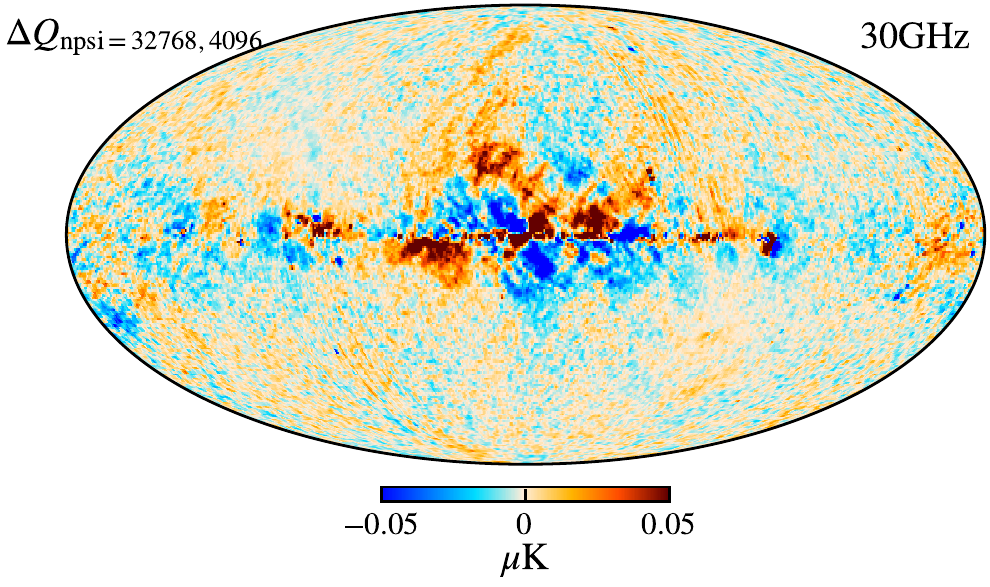}\\
  \includegraphics[width=\linewidth]{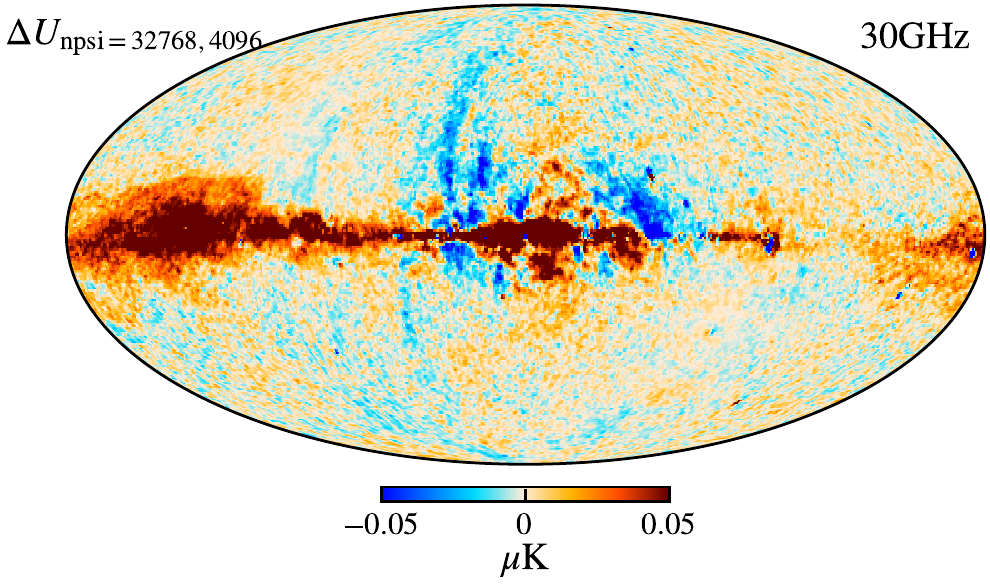}
  \caption{30 GHz T, Q and U map differences between two pipeline executions using different levels of compression, smoothed by a 1 degree beam. The differences in temperature look like correlated noise, and in polarization we have some leakage from polarized synchrotron. In both cases, the amplitude of the differences is much lower than the uncertainties from other effects. 
  }\label{fig:compressiondiff}
\end{figure}

Finally, template components are defined in terms of a user-specified
fixed template map for which the only free parameter is an overall
amplitude at each frequency. This is primarily included for historical
reasons, for instance to support template fitting as implemented by
the \WMAP\ team \citep{bennett2012}. However, this approach allows
very limited uncertainty propagation, and we therefore generally
rather prefer to include relevant survey maps (for instance the
408\,MHz survey by \citealp{haslam1982}) as additional frequency
channels, for which meaningful uncertainties per pixel may be
defined. This mode is not used in the current \BP\ analysis
\citep{BP01}.

The purple boxes in Fig.~\ref{fig:commander} contain sampling routines
for these parameters, and these are split into two categories; linear
and nonlinear. The linear parameters (i.e., component amplitude
parameters) are sampled using a preconditioned Conjugate Gradient
solver \citep{seljebotn:2019}, as implemented in the
\texttt{comm\_cr\_mod} (``Constrained Realization'') module, while the
nonlinear spectral parameters are sampled in the
\texttt{comm\_nonlin\_mod} module using a combination of Metropolis
and inversion samplers \citep{BP13,BP14}.

\section{Optimization}
\label{sec:tricks}

The \commander\ walkthrough given in Sect.~\ref{sec:commander3} is
high-level in nature, and is intended to give a broad overview of the
code. In this section, we turn our attention to lower-level details,
and focus in particular on specific optimization challenges and tricks
that improve computational efficiency.

\subsection{In-memory data compression}
\label{sec:compression}

\begin{table*}[t]
  \begingroup
  \newdimen\tblskip \tblskip=5pt
  \caption{Huffman compression performance for each \Planck\ LFI data
    object and frequency channel. Columns 2--6 are all given in units
    of gigabytes. The last column shows the average ratio between the
    raw and compressed data volumes.}
  \label{tab:compression}
  \nointerlineskip
  \vskip -3mm
  \footnotesize
  \setbox\tablebox=\vbox{
    \newdimen\digitwidth
    \setbox0=\hbox{\rm 0}
    \digitwidth=\wd0
    \catcode`*=\active
    \def*{\kern\digitwidth}
    \newdimen\signwidth
    \setbox0=\hbox{-}
    \signwidth=\wd0
    \catcode`!=\active
    \def!{\kern\signwidth}
 \halign{
      \hbox to 2.5cm{#\leaderfil}\tabskip 1em&
      \hfil#\hfil\tabskip 1em&
      \hfil#\hfil\tabskip 1em&
      \hfil#\hfil\tabskip 1em&
      \hfil#\hfil\tabskip 1em&
      \hfil#\hfil\tabskip 1em&
      \hfil#\hfil\tabskip 1em&      
      \hfil#\hfil\tabskip 0pt\cr
    \noalign{\doubleline}
    \omit&\multispan2\hfil\sc 30\,GHz\hfil&\multispan2\hfil\sc
    44\,GHz\hfil&\multispan2\hfil\sc 70\,GHz\hfil&\omit\cr
    \noalign{\vskip -3pt}
    \omit&\multispan2\hrulefill&\multispan2\hrulefill&\multispan2\hrulefill&\omit\cr
    \noalign{\vskip 3pt}
    \omit\sc Item\hfil&\hfil\sc Raw\hfil&\hfil\sc Huffman
    \hfil&\hfil\sc Raw\hfil&\hfil\sc Huffman\hfil&\hfil\sc Raw
    \hfil&\hfil\sc Huffman\hfil&\hfil\sc Raw/Huffman\hfil\cr
    \noalign{\vskip 4pt\hrule\vskip 4pt}
TOD          & 361 & *52 & *776 & *95 & 2625 & 340 & *6\cr
Pointing ($\hat{n}$) & 181 & *10 & *388 & *18 & 1312 & *69 & 20\cr
Pointing ($\psi$)          & *90 & **5 & *194 & *10 & *656 & *24 & 25\cr
Quality flag         & *45 & **3 & **97 & **6 & *328 & *10 & 33\cr
    \noalign{\vskip 4pt\hrule\vskip 4pt}
Total        & 730 & *70 & 1230 & 130 & 4530 & 450 & 10\cr
    \noalign{\vskip 4pt\hrule\vskip 5pt} } }  
  \endPlancktablewide \endgroup
\end{table*}

As discussed in Sect.~\ref{sec:mem_layout}, one of the key challenges
for achieving efficient end-to-end CMB analysis is memory management;
since the full TOD are required at every single Gibbs iteration, it is
imperative to optimize data access rates. In that respect, we first
note that RAM read rates are typically at least one order of magnitude
higher than disk read rates, and, second, most TOD operations are
bandwidth limited simply because each sample is only used once (or a
few times) per read. For this reason, a very useful optimization step
is simply to be able to store the full TOD in RAM. At the
same time, we note that disk space is cheap, while RAM is
expensive. It is therefore also important to minimize the total data
volume.

We address this issue by storing the TOD in a compressed form in RAM,
and decompress each data segment before further processing at
runtime. In the current implementation, we adopt Huffman coding
\citep{Huffman}, and implement the decompression algorithms natively
in the source code; of course, other lossless compression algorithms
can be used, and in the future it may be worth exploring using
different algorithms for different types of objects.

Typically, most current CMB experiments distribute their data, as
recorded by analogue-to-digital converters (ADCs), in the form of
32-bit integers which support over 2 billion different numbers; we
will refer to each distinct integer as a ``symbol'' in the
following. However, the actual dynamic range of any given data segment
only typically spans a few thousand different symbols. Therefore,
simply by choosing a more economic integer precision level, a factor
of three could be gained. Further improvements could be made by
actually counting the frequency of each symbol separately, and assign
short bit strings to frequently occuring symbols, and longer bit
strings to infrequently occuring symbols. Huffman coding is a
practical algorithm that achieves precisely this, and it can be shown
to be the theoretically optimal lossless compression algorithm when
considering each datum separately. To account for correlations, noting
that most CMB TODs are correlated in time, we difference all
datastreams sample-by-sample prior to compression, setting $\bar{d}(i)
= d(i)-d(i-1)$; after this differencing, most data values will be
close to zero.

The actual Huffman encoding relies on a binary tree structure, and
assigns numbers with high frequencies to short codes near the top of
the tree, and infrequent numbers to long codes near the bottom.  As a
practical and real-life example, Fig.~\ref{fig:huffman} shows the top
of the Huffman encoding tree for the arbitrarily selected Operational
Day (OD) 27\,646 for the 30\,GHz data. In this case, 0 represents
about 42\% of the entire dataset after the pairwise differencing
operation. The optimal compression is therefore to represent 0 with a
single bit (which also happens to be 0), and numbers that occur
frequently, like 1 and $-1$, with 4 bit codes (1110 and 1001
respectively). At the bottom of the tree are those numbers which occur
very infrequently. This diagram is obviously truncated, and the full
tree uses codes with lengths of 20 bits to represent the lowest
symbols that occur only once.

The Huffman algorithm requires a finite number of symbols to be
encoded, and therefore performs far better for integers than for
floats. This is intrinsically the case for (ADC-outputted) TOD and
flag information, but not for pointing values. However, as most modern
CMB experiments, \BP\ uses HEALPix to discretize the
sky. Precomputing the HEALPix coordinates of each sample therefore
allows the pointing sky coordinates to be reduced from two floats per
sample (representing $\theta$ and $\phi$) to one single integer. Of
course, this requires that the HEALPix resolution parameter,
$N_{\mathrm{side}}$, is predefined during data preprocessing and
compression for each band. In practice, this is acceptable as long as
$N_{\mathrm{side}}$ is selected to correspond to a higher resolution
than the natural beam smoothing scale of the detectors. For the LFI 30
and 44\,GHz channels, we adopt $N_{\mathrm{side}}=512$, whereas for
the 70\,GHz channel we adopt $N_{\mathrm{side}}=1024$.\footnote{Note
  that this is different from the official \Planck\ LFI maps, which
  adopt $N_{\mathrm{side}}=1024$ for all channels.} We note that this
is indeed a lossy compression step, but it is precisely the same lossy
compression that is always involved in CMB mapmaking; the only
difference is that the discretized pointing is evaluated once as a
preprocessing step.

The same does not hold for the polarization angle, $\psi$, which also
is a float, and typically is not discretized in most current CMB
mapmaking codes. However, as shown by \citet{keihanen2012} in the
context of beam deconvolution for LFI through the use of so-called 3D
maps, even this quantity may be discretized with negligible errors as
long as a sufficient number of bins between 0 and $2\pi$ is
used. Specifically, \citet{keihanen2012} used $n_{\psi}=256$ bins for
all LFI channels, while for \BP\ we adopt $n_{\psi}=4096$; the
additional resolution has an entirely negligible cost in terms of
increased Huffman tree size.

To check the impact of the polarization angle compression,
Fig.~\ref{fig:compressiondiff} shows difference Stokes $T$, $Q$, and $U$
maps for the coadded 30\,GHz frequency channel generated in two
otherwise identical mapmaking runs with $n_{\psi}=4096$ and
32\,768. Here we see that the compression introduces artifacts at the
level of 0.01 $\mu$K at high Galactic latitudes, increasing to about
$0.1\,\mu$K in the Galactic plane, which is entirely negligible
compared to the intrinsic uncertainties in these data, fully
confirming the conclusions of \citet{keihanen2012}.

The main cost associated with Huffman compression comes in the form of
an additional cost for decompression prior to TOD
processing. Specifically, we find that decompression costs about 10\%
of the total runtime per iteration; we consider this to be an
acceptable compromise to enable the entire dataset to be held in
memory at once and eliminate disk read time.

Table~\ref{tab:compression} gives an overview of the compression
performance for each data object and LFI frequency channel. Overall,
we see that the TOD data volume is reduced by a factor of about six,
while the pointing volume is reduced by a factor of about 20. 

\subsection{FFT optimization and aliasing mitigation}

Once the data are stored in memory, the dominant TOD operation is
the Fast Fourier Transforms (FFT), which are used repeatedly for both
correlated noise and gain sampling \citep{BP06,BP07}. Fortunately,
several highly optimized FFT libraries are widely available that
provides outstanding performance, and we currently adopt the FFTW
implementation \citep{FFTW05}.

\begin{figure}[t]
  \center
  \includegraphics[width=\linewidth]{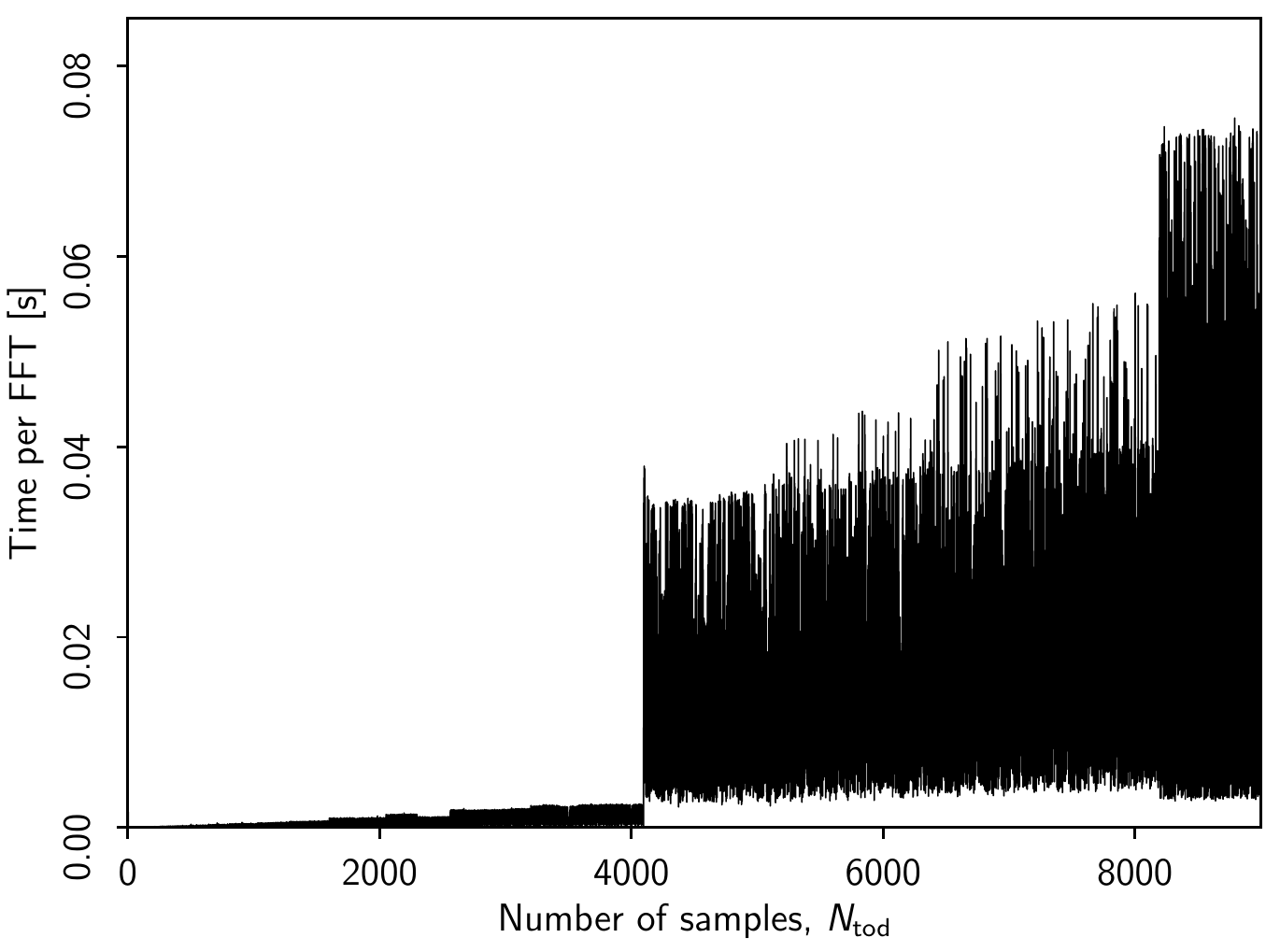}\\
  \includegraphics[width=\linewidth]{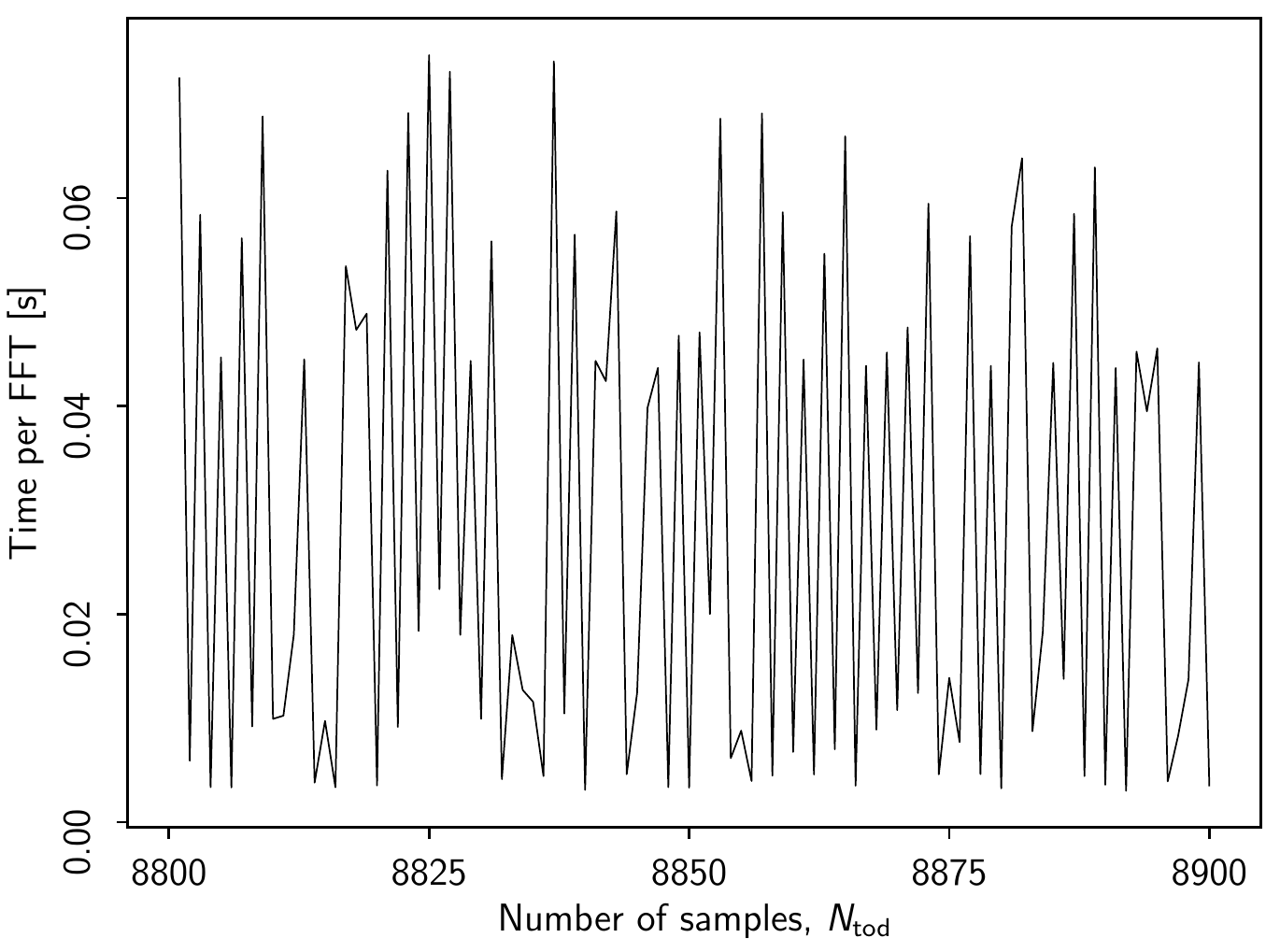}
  \caption{(\emph{Top:}) Cost per FFT as a function of FFT
    length. (\emph{Bottom:}) Zoom-in of the top panel, showing more
    details of the variation within a small sample range.
  }\label{fig:fftw}
\end{figure}

Still, there are several issues that needs to be considered regarding
FFTs. The first regards runtime versus the length of each data
segment. In particular, the FFTW documentation notes that
\begin{displayquote}
``\emph{FFTW works most efficiently for arrays whose size can be factored into small primes (2, 3, 5, and 7).''}
\end{displayquote}
As an illustration of this fact, Fig.~\ref{fig:fftw} shows the time
per FFT as a function of data length, as measured on a local compute
cluster; the bottom panel shows all lengths up to 10\,000, while the
bottom panel shows a zoom-in of the top panel. Here we clearly see
that run times can vary by at least an order of magnitude from one
length to the next.

\commanderthree\ exploits this effect at read-in time. As each chunk
of data is read, its length is compared to a precomputed table of FFT
costs for all values up to $N_{\mathrm{samp}} = 10^6$. The data chunk
is then truncated to the nearest local minimum of the cost function,
losing only a small amount of data (0.03\%) while providing a large
speedup for all FFT operations performed on that chunk. Of course, if
the noise stationarity length is unconstrained, and the segment length
is fully up to the user to decide, then powers of $2^{n}$ are
particularly desirable.

Another important effect to take into account is that of FFT aliasing
and edge effects; the underlying FFT algebra assumes by construction
that the data in question are perfectly periodic. If there are notable
gradients extending through the segment, the end of the segment may
have a significantly different mean than the beginning, and this will
be interpreted by the FFT as a large discrete jump. If any filtering
or convolution operators (for instance inverse noise weighting,
$\N^{-1}$) are applied to the data, this step will result in
non-localized ringing that can contaminate the data.

Several approaches to mitigate this effect are described in the
literature, and zero padding is perhaps one of the best known, in
which one adds zeros to both ends of the original data
stream. However, this operation also has nontrivial impact on the
outcome, and we choose a more expensive and more conservative
approach, namely mirroring: Before every FFT operation, we double the
length of the array in question, and add a mirrored version of the
original array into the second half. This makes the function periodic
by construction, and eliminates any discrete boundary
effects. However, this safety does come at a price of doubling the run
time, and future implementations should explore alternative
approaches. One possibility is to allow for data duplication in
overlap regions, such that for instance 5\,\% of a given data segment
is filled by data from the neighboring segments at both edges. Then
the overlap region is discarded after filtering. This approach was
explored by Galloway (2018),\footnote{PhD thesis.} and shown to work
very well for SPIDER noise modelling.

\subsection{Conjugate Gradient optimization for component separation}
\label{sec:cg}

The second most important numerical operation in the \BP\ Gibbs
sampler after FFTs is the spherical harmonics transform. This forms
the numerical basis for the astrophysical component amplitude sampler
\citep{BP13}, in which the following equation is solved repeatedly
\citep{seljebotn:2019},
\begin{equation}
\biggl(\S^{-1} +
\sum_{\nu}\Y^t_{\nu}\M^t_{\nu}\N_{\nu}^{-1}\M_{\nu}\Y_{\nu}\biggr)\,\a
= \sum_\nu\Y_{\nu}^t\M_\nu^t\N_\nu^{-1}\m_{\nu}.
\label{eq:wiener}
\end{equation}
Here $\S$ and $\N_\nu$ denote the signal and noise covariance
matrices, respectively, $\M_{\nu}$ is a mixing matrix, $\m_{\nu}$ is
an observed frequency map, $\a$ is a vector containing all component
amplitudes, and $\Y$ is a spherical harmonics transform. In this
expression, $\N_{\nu}$, $\M_{\nu}$, and $\m_{\nu}$ are all defined as
pixelized map vectors, while $\S$ and $\a$ are defined in spherical
harmonic space, and $\Y$ converts between the two spaces.

Equation~\eqref{eq:wiener} involves millions of free parameters, and
must therefore be solved iteratively with preconditioned Conjugate
Gradient (CG) type methods \citep{shewchuk:1994}. There are therefore
two main approaches to speed up its solution: Either one may reduce
the computational cost per CG iteration, or one may reduce the number
of iterations required for convergence. As far as the former approach
is concerned, by far the most important point is simply to use the
most efficient SHT library available at any given time to perform the
$\Y$ operation; all other operations are linear in the number of
pixels or spherical harmonics coefficients, and are largely irrelevant
as far as computational costs are concerned. At the time of writing,
the fastest publicly avaialble SHT library is \texttt{libsharp2}
\citep{reinecke2013}, and we employ its MPI version for the current
calculations. (We note that the OpenMP version is even faster, but
since the current \commanderthree\ parallelization strategy is
agnostic with respect to compute nodes, and all data are parallelized
across all available nodes, this mode is not yet supported.)

The main issue to optimize is therefore the number of iterations
required for convergence. Again, there are two different aspects to
consider, namely preconditioning and the stopping criterion. Starting
with the former, we recall that a preconditioner is simply some
(positive definite) linear operator, $\P$, that is applied to both
sides of Eq.~\eqref{eq:wiener} in the hope that the equation becomes
easier to solve numerically. The ideal case is that $\P$ is equal to
the inverse of the coefficient matrix on the left-hand side, but this
is of course never readily available; if it were, the system would
already be solved. In the current work, we adopt the preconditioner
introduced by \citet{seljebotn:2019} for Eq.~\eqref{eq:wiener}, which
approximates the inverse of a non-square matrix, $\A$, by its
pseudo-inverse $\A^{+}\equiv(\A^t\A)^{-1}\A^t$. A possible future
improvement might be to replace this on large angular scales with the
exact brute-force block preconditioner of
\citet{eriksen:2004,eriksen2008} for $\ell\lesssim 100$.

\begin{figure}[t]
  \center
  \includegraphics[width=\linewidth]{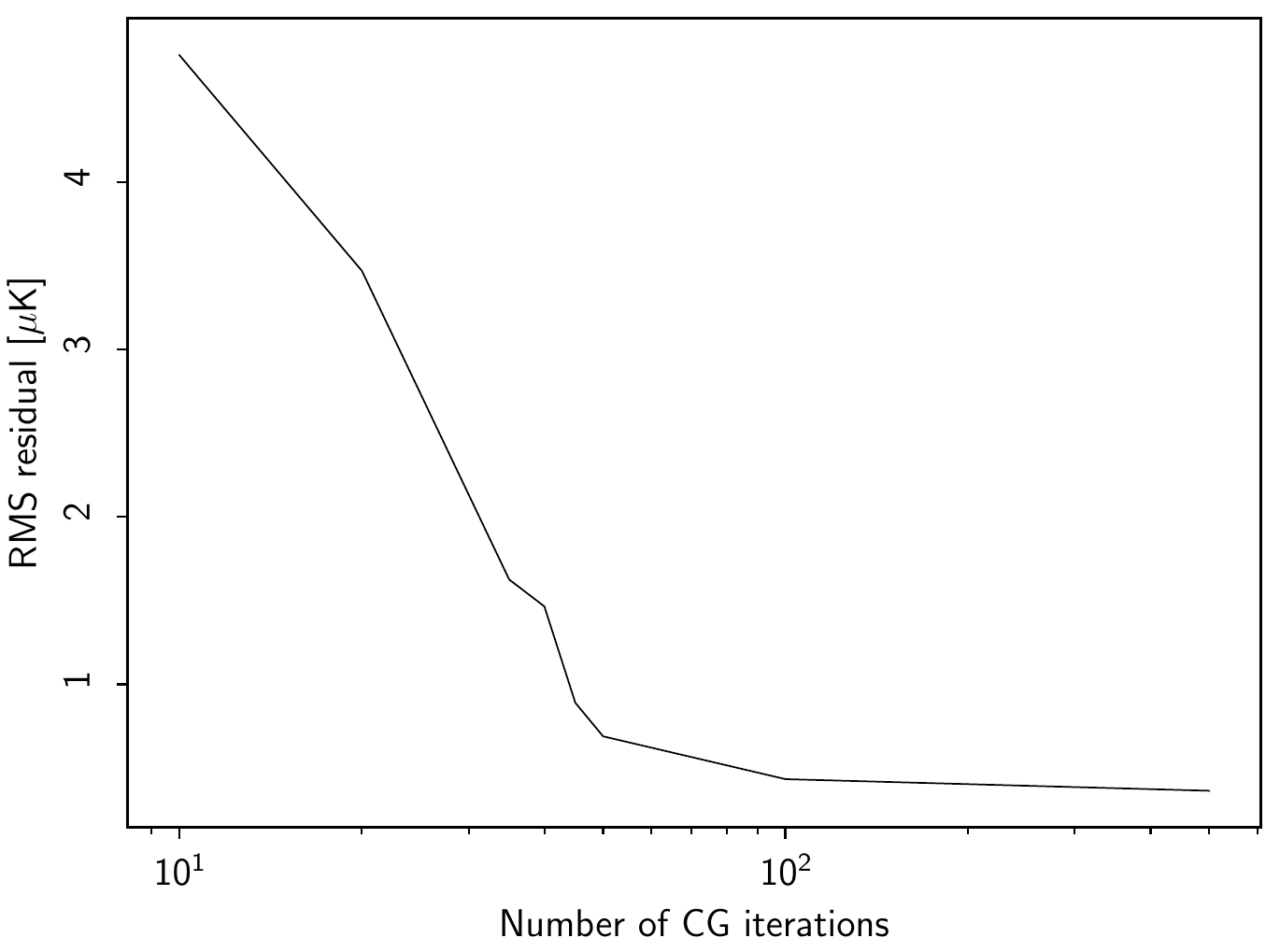}
  \caption{RMS of a CMB difference map comparing various iteration numbers to the most converged 1000 iteration case. The RMS drops rapidly until about 50 samples, at which point the marginal increase in convergence per sample flattens out.
  }\label{fig:rmsiterations}
\end{figure}

The final question is then, simply, to determine how many CG
iterations are required to achieve acceptable accuracy. To address
this issue, we solve Eq.~\eqref{eq:wiener} for the basic
\BP\ configuration \citep{BP01} with a maximum of 1000 iterations, and
plot the rms difference between the CMB solutions obtained at the
$i$th and 1000th iterations. This quantity is plotted in
Fig.~\ref{fig:rmsiterations}. Here we see that that the residual
decreases rapidly up to about 70 iterations, while for more than 100
iterations only very modest differences are seen. For the final
\BP\ runs, we have chosen 100 iterations as the final cut-off.

We note that this criterion differs from most previous
\commander-based analyses \citep[e.g.,][]{planck2014-a12}, which
usually have defined the cut-off in terms of a relative reduction of
the preconditioned residual, $r = ||\A\x-\b)|^2$. The reason we prefer
to define the convergence criterion in terms of map-level residuals
with respect to the converged solution is simply that $\r$ may weight
the various astrophysical components very differently, for instance
according to arbitrarily chosen units. One example is synchrotron
emission, which has a reference frequency of 408\,MHz in the
\BP\ analysis, and is measured in units of
$\mu\textrm{K}_{\mathrm{RJ}}$, and therefore has a much higher impact
on $r$ than the CMB component. If we were to use the preconditioned
residual as a threshold instead (which many analyses also do), then
nearly-singular modes in $\A$ may be given a relatively large
weight. In practice, it is our experience that a map-based rms cut-off
is less prone to spurious and premature termination than either of the two
residual-based criteria.

\subsection{File format comparison; HDF versus FITS}

\begin{figure}[t]
  \center
  \includegraphics[width=\linewidth]{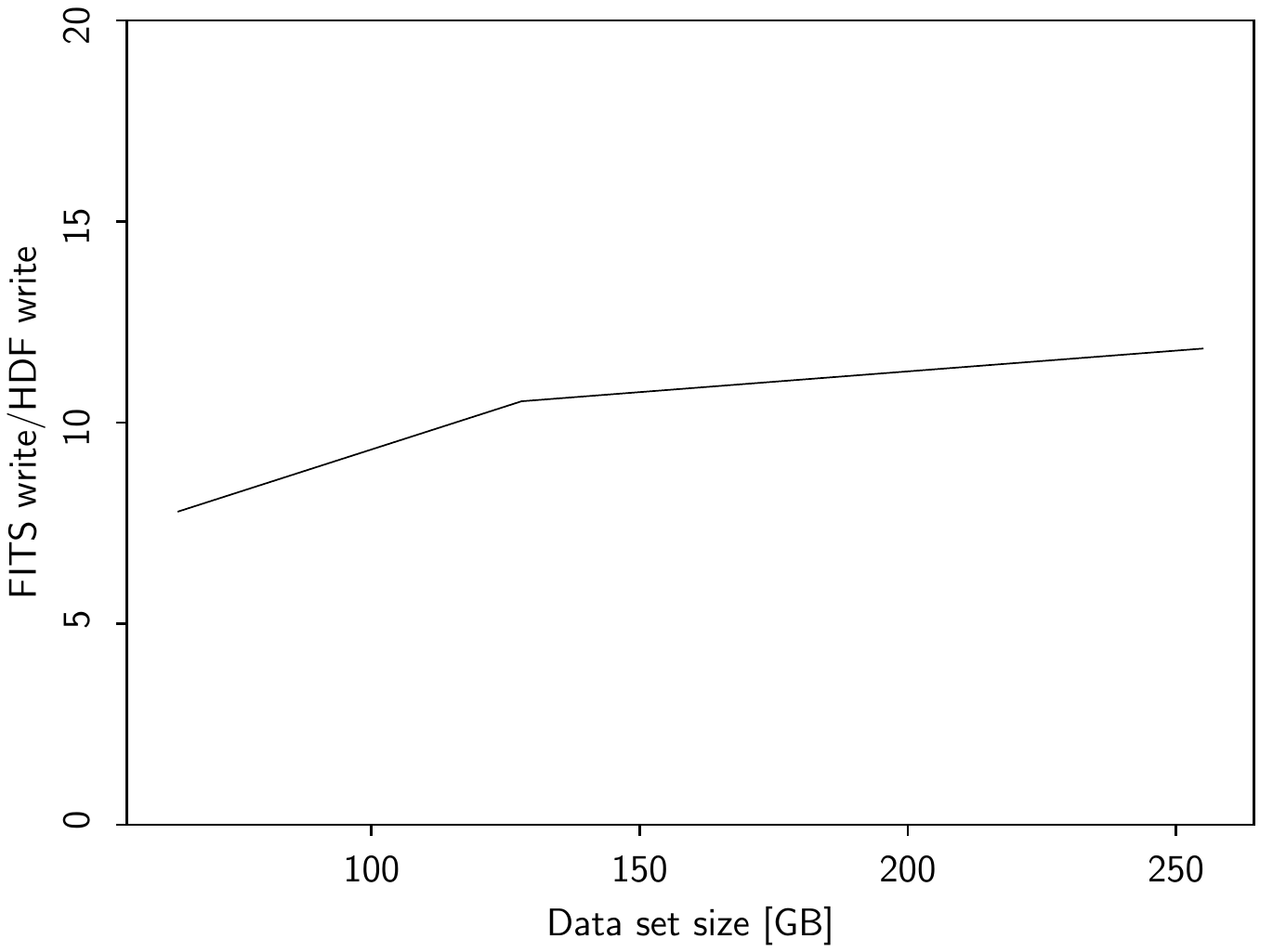}
  \caption{Ratio between FITS and HDF disk write times for subsets of
    the LFI data of various sizes. 
  }\label{fig:hdfVfits}
\end{figure}

\begin{table*}[t]
  \begingroup
  \newdimen\tblskip \tblskip=5pt
  \caption{Computational resources required for end-to-end
    \BP\ processing. All times correspond to CPU hours. All reported times are
    averaged over more than 100 samples, and vary by $\lesssim\,5\,\%$ from sample to
    sample.}
  \label{tab:resources}
  \nointerlineskip
  \vskip -3mm
  \footnotesize
  \setbox\tablebox=\vbox{
    \newdimen\digitwidth
    \setbox0=\hbox{\rm 0}
    \digitwidth=\wd0
    \catcode`*=\active
    \def*{\kern\digitwidth}
    \newdimen\signwidth
    \setbox0=\hbox{-}
    \signwidth=\wd0
    \catcode`!=\active
    \def!{\kern\signwidth}
 \halign{
      \hbox to 7.5cm{#\leaderfil}\tabskip 1em&
      \hfil#\tabskip 1em&
      \hfil#\tabskip 1em&
      \hfil#\tabskip 1em&
      \hfil#\tabskip 2em&
      #\tabskip 0em\hfil\cr
    \noalign{\doubleline}
      \omit\textsc{Item}\hfil&
      \omit\hfil\textsc{30 GHz}\hfil&
      \omit\hfil\textsc{44 GHz}\hfil&
      \omit\hfil\textsc{70 GHz}\hfil&
      \omit\hfil\textsc{Sum}\hfil&
      \omit\hfil\textsc{Reference}\hfil\cr
      \noalign{\vskip 4pt\hrule\vskip 4pt}
      %\noalign{\vskip 8pt}
      %\noalign{\vskip 5pt\hrule\vskip 5pt}
      \multispan5\textit{Data volume}\hfil\cr
      \noalign{\vskip 2pt}
      \hskip 10pt Uncompressed TOD volume                 & {\gray 761 GB} &
      {\gray 1\,633 GB} & {\gray 5\,522 GB} & {\gray 7\,915 GB}& \cr
      \hskip 10pt Compressed TOD volume & **86
      GB & *178 GB & **597 GB & ***861 GB& \cr
      \hskip 10pt Non-TOD-related RAM usage &   &  &  & ***659 GB& \cr
      \hskip 10pt {\bf Total RAM requirements} &  &  &  & **{\bf1\,520 GB}& \cr      
      \noalign{\vskip 2pt}      
      \multispan5\textit{Processing time (cost per run)}\hfil\cr
      \hskip 10pt TOD initialization/IO time                    & 3.8\,h & 4.3\,h & 12.5\,h & 20.6\,h& \cr
      \hskip 10pt Other initialization                          &  &  &  &  43.4\,h& \cr
      \hskip 10pt {\bf Total initialization}                          &  &  &  &  {\bf 64.0\,h}& \cr
      \noalign{\vskip 2pt}      
      \multispan5\textit{Gibbs sampling steps (cost per sample)}\hfil\cr
      %\noalign{\vskip 2pt}
      \hskip 10pt Huffman decompression                            & 1.1\,h& 1.8\,h& 7.1\,h& 10.0\,h& This paper\cr
      \hskip 10pt TOD projection ($\P$ operation)               & 0.3\,h& 0.7\,h& 3.1\,h&  4.1\,h& \citet{BP01}\cr
      \hskip 10pt Sidelobe evaluation ($\s_{\mathrm{sl}}$)         & 1.1\,h& 2.1\,h& 6.5\,h&  9.7\,h& \citet{BP08}\cr
      \hskip 10pt Orbital dipole ($\s_{\mathrm{orb}}$)             & 0.5\,h& 1.1\,h& 4.6\,h& 6.2\,h& \citet{BP07}\cr
      \hskip 10pt Gain sampling ($g$)                           & 0.6\,h& 0.7\,h& 4.7\,h& 6.0\,h& \citet{BP07}\cr
      \hskip 10pt 1\,Hz spike sampling ($s_{\mathrm{1hz}}$)      &
      0.2\,h& 0.3\,h& 1.9\,h& 2.4\,h& \citet{BP01}\cr      
      \hskip 10pt Correlated noise sampling ($\n_{\mathrm{corr}}$) & 1.7\,h& 3.6\,h& 24.8\,h& 30.1\,h& \citet{BP06}\cr
      \hskip 10pt Correlated noise PSD sampling ($\xi_{\mathrm{n}}$) & 3.3\,h& 4.0\,h& 1.1\,h& 8.4\,h& \citet{BP06}\cr
      \hskip 10pt TOD binning ($\P^t$ operation)                &
      0.2\,h& 0.5\,h& 4.1\,h& 4.8\,h& \citet{BP10}\cr
     %\hskip 10pt Map solution ($Ax=b$ for each pixel)          & *13sec& *22 sec& *128 sec& *163 sec\cr
      \hskip 10pt Sum of other TOD processing                   & 1.3\,h& 2.5\,h& 10.9\,h& 14.7\,h& \citet{BP01}\cr
      \hskip 10pt {\bf TOD processing cost per sample}          & {\bf
        10.4\,h}& {\bf 17.4\,h}& {\bf 69.1\,h}&  {\bf 96.9\,h}& \cr
      \noalign{\vskip 2pt}
      \hskip 10pt Amplitude sampling, $P(\a\mid \d, \omega\setminus\a)$  &   &  &  & 23.9\,h& \citet{BP13}\cr
      \hskip 10pt Spectral index sampling, $P(\beta\mid \d, \omega\setminus\beta)$  &   &  &  & 40.3\,h& \citet{BP14}\cr
      \hskip 10pt Other steps                                   &      &  &  &  0.6\,h& \citet{BP01}\cr
      \noalign{\vskip 2pt}
      \hskip 10pt {\bf Total cost per sample}                   &   &  &  &  {\bf 163.9\,h}& \cr
      \noalign{\vskip 4pt\hrule\vskip 5pt} } }
  \endPlancktablewide \endgroup
\end{table*}

The current \commander3\ implementation adopts the Hierarchical Data
Format (HDF) for TOD disk storage. While the CMB community has largely
converged on the standard FITS format for maps, this format has some
drawbacks which make them less than ideal for time-domain datasets,
both in terms of efficiency and programming convenience. For instance,
HDF files support internal directory tree structures, which allows for
intuitive storage of multiple layers of information within each
file. Additionally, HDF can easily support datasets with different
lengths, which is useful when handling compressed data. Finally, the
HDF format supports headers and metadata for every dataset, which
makes it very easy to store quantities such as units, conversion
factors, compression information and even human-readable help strings
locally.

Most importantly, however, is simply the fact that HDF is faster than
FITS. To quantify this, we have performed several timing tests, and
one example is shown in Fig.~\ref{fig:hdfVfits}. In this case, we
write a given data subset of varying size (single-detector 27M,
single-horn 27M+S and the full 30\,GHz channel) to disk repeatedly
using standard Python libraries, and we plot the ratio of the time
averages required for this task. HDF operations are performed with
\texttt{h5py} and FITS operations with \texttt{astropy.io.fits}. For
this particular case, we see that HDF output is typically one order of
magnitude faster than FITS output on our system.

\section{Resource requirements}

At the outset of the \BP\ project, it was by no means obvious whether
full end-to-end Bayesian processing was computationally feasible with
currently available computing resources. A main goal of the general
project in general, and this paper in particular, was therefore simply
to quantify the resource requirements for end-to-end Bayesian CMB
analysis in a real-life setting, both in terms of CPU hours and
RAM. These are summarized for the main \BP\ run, as defined in
\citet{BP01}, in Table~\ref{tab:resources}. All processing times refer
to total CPU hours integrated over computing cores, and since each
chain is parallelized over 128 cores, all numbers may be divided by
that number to obtain wall hours.

Starting with the data volume, we see that the total raw LFI TOD span
almost 8\,TB as provided by the \Planck\ Data Processing Center
(DPC). After compression and removing
non-essential information, this is reduced by almost an order of
magnitude, as the final RAM requirements for LFI TOD storage is only
861\,GB. The total RAM requirement for the full job including
component separation (most of which is spent on storing the full set
of mixing matrices per astrophysical component and detector) is
1.5\,TB.

The second section shows the total initialization time, which accounts
for the one-time cost of reading all data into memory. This is
mostly dominated by disk read times, so systems with faster disks will
see improvements here. However, as this is only executed at the start
of the run, it is a very subdominant cost compared to the loop
execution time.

The bottom section of Table~\ref{tab:resources} summarizes the
computational costs for a single iteration of the Gibbs sampling
loop. Here we see that the total cost per sample is dominated by the
TOD sampling loop (as would be naively expected from data volume),
which takes about 59\,\% of the full sample time. The remaining 41\,\%
is spent on component separation, and about one third of this is spent
on amplitude sampling (as discussed in Sect.~\ref{sec:cg}), and two
thirds is spent on spectral parameter sampling. The former of these is
fairly well optimized, as it is dominated by SHT's, while the latter
clearly could be better optimized, for a potential maximum saving of
25\,\% of the total runtime. 

For the 70\,GHz channel, 35\,\% of the total processing time is spent
on correlated noise sampling, $P(\n_{\mathrm{corr}}\mid\d,\ldots)$. This
step is by itself the most computationally complex and expensive
operation, as it requires FFTs within a CG solver for correlated noise
gap filling \citep{BP06}. This step could be sped up through
approximations, but we found that the CG solver was the only way to
guarantee high accuracy in regions with large processing masks, most
notably scans through the Galactic plane. For the 30 and 44\,GHz
channels, the most expensive operation is in fact correlated noise PSD
sampling, $P(\xi_{\mathrm{corr}}\mid\d,\ldots)$, and this is an
indication of sub-optimality of the current implementation, rather
than a fundamental algorithmic bottleneck: One of the last
modifications made to the final \BP\ pipeline was the inclusion of a
Gaussian peak in the noise PSD around 1\,Hz, and this operation was
not optimized before the final production run. Future implementations
should be able to reduce this time to negligible levels, as only
low-volume power spectrum data are involved in noise PSD sampling.

Next, we see that the Huffman decompression costs about 10\,\% of the
total runtime, and we consider this to be a fair price to pay for
significantly reduced memory requirements. Indeed, without Huffman
compression we would require multi-node MPI communication, and in that
case many other operations would become significantly more
expensive. Thus, it is very likely that Huffman coding in fact leads
to both lower memory requirements and reduced total runtime.

We also see that sidelobe evaluation accounts for about 10\,\% of the
total runtime, most of which is spent on interpolation. This part can
also very likely be significantly optimized, and the
\texttt{ducc}\footnote{\url{https://gitlab.mpcdf.mpg.de/mtr/ducc}}
library appears to be a particularly promising candidate for future
integration. Sidelobe evaluation will become even more important for a
future \WMAP\ analysis, for which four distinct detector TODs are
combined into a single differencing assembly TOD prior to noise
estimation and mapmaking, each with its own bandpass
\citep{bennett2012}. Actually, as reported by \citet{BP17}, sidelobe
interpolation currently accounts for about 20\,\% of the total
\WMAP\ runtime due to this structure.

Unaccounted TOD processing steps represent a total of 12\,\% of the
total low-level processing time, and include both actual computations,
such as $\chi^2$ evaluations, bandpass sampling etc., but also loss
due to poor load balancing. The latter could clearly be reduced in
future versions of the code.

Overall, we see from Table~\ref{tab:resources} that up to 74\,h per
sample (sidelobe evaluation, correlated noise PSD sampling, spectral
index sampling, and other TOD costs) can potentially be gained through
more careful optimization within the current coding paradigm, or about
half of the total runtime. This optimization will of course happen
naturally in the future, as each module gradually matures. Also, the
current native \commanderthree\ code does not yet support
vectorization (SSE, AVX, etc.) natively, but only partially through
the external FFT and SHT libraries, and whatever the Fortran compiler
can manage on its own. This will also be done in future work.

Finally, it is important to note that the current analysis framework
is inherently a Markov chain, and that means that each sample depends
directly on the previous one. The Bayesian analysis approach is
therefore intrinsically more challenging to parallelize than the
forward simulation frequentist-style approach, for which independent
realizations may be run on separate compute cores
\citep[e.g.,][]{planck2014-a14}. As each variable in the Gibbs chain
is required to hold the full previous state of the chain constant, it
is difficult to find segments of the code that can run independently
for long times without synchronizing. One place this could be added is
in the TOD sampling step. Each of the different bands could be run
independently at the same time (i.e., the 30\,GHz map is independent
of the 44 and 70\,GHz parameters). With many TOD bands, like the
configuration proposed for LiteBIRD, this could be a feasible
parallelization scheme, but for LFI with only three bands the runtime
is dominated by the 70 GHz processing time regardless, and this
technique could shave at most 20\,\% off the total runtime for the
current run. For future analyses of massive data sets with thousands
of detectors, however, data partitioning will become essential to
achieve acceptable parallelization speedup.

\section{Summary and outlook}
\label{sec:conclusions}

The main goal of this paper is to provide an overview of the
\BP\ infrastructure used to analyze the \Planck\ LFI data within an
end-to-end Bayesian framework, hopefully aiding new users to modify
and extend it to their needs. We have discussed the various
computational and architectural decisions that have been adopted for
the \BP\ codebase, as well as some of the current challenges facing
the development effort. We highlight in particular the choices made
for the analysis of the LFI data, but many of these architectural
decisions were selected to be generalizable to future datasets.

One important novel feature introduced here that is likely to be
useful for many future CMB experiments is in-memory data
compression. We find that the original data volume may be reduced by
one order of magnitude through data selection and compression, with
negligible loss of precision. For LFI, this allows the entire dataset
to be stored in memory on modest computing hardware, and it reduces
the disk read time to a one-time initialization cost, independent of
the number of iterations of the algorithm. In general, in-memory
compression permits the analysis to be performed on small clusters
that are often available at individual research institutions, as
opposed to national high-performance computing centers, and this has
significant advantages in both cost and ease of use, for instance
shorter debugging cycles and queuing times. We suggest that future
experiments such as CMB-S4 and Simons Observatory that are planning for data
volumes many times larger than \Planck's consider using these lossless
techniques to reduce the resource requirements of their overall
analysis task.

We also quantify the computational costs for the \BP\ LFI analysis,
and the resulting numbers may serve as an estimate of the pipeline's
performance for future similarly sized datasets. Specifically, we find
that the LFI analysis requires 1.5\,TB of RAM, and producing one
single sample costs about 170 CPU-hrs. This latter number may be
compared with the costs required to produce the official \Planck\ Full
Focal Plane simulations. For instance, as discussed by
\citet{planck2014-a14}, producing 81\,000 LFI noise simulations on the
Finnish \emph{Sisu} cluster cost 4 million CPU hours, for an average
cost of 50\,CPU-hrs/map. The current \BP\ approach, which includes all
steps from low-level calibration to final component separation and
allows for full exploration of parameter degeneracies, is therefore
computationally equivalent to producing only three
correlated-plus-white noise realizations in a traditional frequentist
approach, which offers significantly less powerful error propagation. We
conclude that the Bayesian approach compares favorably with respect to
the traditional approach in terms of computational costs.

More generally, we conclude that the analysis pipeline described in
this paper is ideally suited for moderately-sized CMB datasets, and we
believe that it can be extended to many existing and future
experiments with relatively minor efforts. One concrete and specific
example is the on-going \WMAP\ analysis presented by \citet{BP17},
which appears quite encouraging both in terms of computational
efficiency and data quality.

While the \BP\ project itself was a time-limited effort from 2018 to
2021, this work will be continued within the context of the Open
Science and community-wide \textsc{Cosmoglobe} project. We strongly
encourage all interested parties to get involved in that project, and
together develop an Open Source state-of-the-art model of CMB sky.

\begin{acknowledgements}
  We thank Prof.\ Pedro Ferreira and Dr.\ Charles Lawrence for useful suggestions, comments and 
  discussions. We also thank the entire \Planck\ and \WMAP\ teams for
  invaluable support and discussions, and for their dedicated efforts
  through several decades without which this work would not be
  possible. The current work has received funding from the European
  Union’s Horizon 2020 research and innovation programme under grant
  agreement numbers 776282 (COMPET-4; \BP), 772253 (ERC;
  \textsc{bits2cosmology}), and 819478 (ERC; \textsc{Cosmoglobe}). In
  addition, the collaboration acknowledges support from ESA; ASI and
  INAF (Italy); NASA and DoE (USA); Tekes, Academy of Finland (grant
   no.\ 295113), CSC, and Magnus Ehrnrooth foundation (Finland); RCN
  (Norway; grant nos.\ 263011, 274990); and PRACE (EU).
\end{acknowledgements}

\bibliographystyle{aa}

\bibliography{Planck_bib,BP_bibliography,sources.bib}

\end{document}